\documentclass[aps,prd,twocolumn,superscriptaddress,showpacs]{revtex4}
\usepackage{epsfig,epsf}
\usepackage{amsmath}
\usepackage{amsthm}
\usepackage{amsfonts}
\usepackage{amssymb}
\usepackage{dsfont}
\usepackage{multirow}
\usepackage{appendix}
\usepackage{slashed}
\usepackage[active]{srcltx}
\usepackage{psfrag}

\begin{document}

\title{ Properties of $\Sigma_Q^{*}$, $\Xi_Q^{*}$ and   $\Omega_Q^{*}$ heavy baryons in cold nuclear matter}
\date{\today}
\author{K.~Azizi}
\affiliation{School of Physics, Institute for Research in Fundamental Sciences (IPM), P. O. Box 19395-5531, Tehran, Iran}
\affiliation{Department of Physics, Do\v{g}u\c{s} University, Acibadem-Kadik\"{o}y, 34722
Istanbul, Turkey}
\author{N.~Er}
\affiliation{Department of Physics, Abant \.{I}zzet Baysal University,
G\"olk\"oy Kamp\"us\"u, 14980 Bolu, Turkey}

\begin{abstract}
The in-medium properties of the  heavy spin-3/2 $\Sigma_Q^{*}$, $\Xi_Q^{*}$ and  $\Omega_Q^{*}$ baryons with $Q$ being $b$ or $c$ quark are investigated. The shifts in some spectroscopic parameters of these particles due to the saturated cold nuclear matter are calculated. The variations of those parameters with respect to the changes in the density of the cold nuclear medium are studied, as well.  It is observed that the parameters of  $\Sigma_Q^{*}$ baryons are considerably affected by the  nuclear matter compared to the $\Xi_Q^{*}$ and  $\Omega_Q^{*}$ particles that roughly do not see the medium. The results obtained may be used in analyses of the data to be provided by the in-medium experiments like PANDA.
\end{abstract}


\maketitle

\section{Introduction}
The vacuum and in-medium hadronic spectroscopies are powerful tools for investigation and understanding of the internal structures of hadrons as well as the perturbative and non-perturbative natures of QCD. In the last two decades, the spectroscopy of heavy baryons has been received much attention both theoretically and experimentally. All  heavy baryons with a single heavy $b$ or $c$ quarks predicted by the quark model have been discovered \cite{PDG}, except $\Omega_b^*$ state that is expected to be found in near future considering the new developments in experimental side. Indeed, many ground state and excited resonances in different channels have also been discovered  \cite{Artuso:2000xy,Aubert:2007dt,Lesiak:2008wz,Aaltonen:2011sf,Aaij:2016jnn,Aubert:2006je,Solovieva:2008fw,Aaltonen:2007ar}.  Very recently, the LHCb Collaboration found five narrow resonances, $\Omega_c(3000), \Omega_c(3050), \Omega_c(3066), \Omega_c(3090), \Omega_c(3119)$ in $\Omega_c$ channel in the $\Xi_c^+ K^-$ invariant mass spectrum \cite{Aaij:2017nav}, but unfortunately did not fix their quantum numbers.  After the discovery, many theoretical manipulations on the nature and structure of these states were appeared \cite{Agaev:2017jyt,Agaev:2017lip,Karliner:2017kfm,Wang:2017vnc,Chen:2017gnu,Aliev:2017led,Yang:2017rpg,Huang:2017dwn,Zhao:2017fov,Padmanath:2017lng}. Some authors treated them as usual three-quark resonances \cite{Agaev:2017jyt,Agaev:2017lip,Karliner:2017kfm,Wang:2017vnc}, some other interpreted them as  new penta-quark states \cite{Yang:2017rpg,Huang:2017dwn}. Experimental and theoretical works on the nature of these new resonances are continued. In the doubly heavy baryons' side, the doubly charmed $\Xi^{++}_{cc}$ was newly founded in $\Lambda_c^+K^-\pi^+\pi^-$ mass spectrum by LHCb Collaboration \cite{Aaij:2017ueg}. In the exotic sector many heavy tetraquark and pentaquark structures have been discovered, as well \cite{Ablikim:2015tbp,Ablikim:2013mio,Choi:2003ue,Choi:2007wga,Mizuk:2009da,Chilikin:2013tch,Belle:2011aa,Garmash:2014dhx, Liu:2013dau,Aaij:2016iza,Aaij:2016nsc}. By the newly recorded progresses, we hope that we will found more exotic resonances and complete the picture of the doubly and triply charmed/bottom baryons previously predicted by the theory  \cite{Agaev:2017uky,Azizi:2016dhy,Agaev:2016dsg,Azizi:2017bgs,Agaev:2017oay,Aliev:2012iv,Aliev:2014lxa,Aliev:2012tt}. 

The single heavy baryons have already been studied in vacuum within different  non-perturbative approaches, like the relativistic quark model, chiral quark model, QCD sum rules, etc  \cite{Azizi:2015ksa,Chen:2015kpa,Chen:2016phw,Wang:2008hz,Wang:2007sqa,Wang:2009cr,Ebert:1996ec,Liu:2012sj,Aliev:2008sk}. To better analyze the results of the in-medium and heavy-ion collision experiments as well as the heavy baryon-nucleon interaction, we need to study the in medium properties of these baryons, as well. The study of many parameters of hadrons in nuclear matter is  an important tool for better understanding of the hadronic dynamics under extreme conditions. Such investigations can also help us get deeper knowledge on the nature of the quark-gluon plasma as a new phase of matter. The in-medium properties of many hadrons have been already investigated (for instance see \cite{Liang:2013sqa,Jeong:2016qlk,Jin:1994bh,Wang:2011hta,Wang:2012xk,Drukarev:2013kga,Azizi:2016dmr,Azizi:2016hbr,Hosaka:2016ypm} and the references therein). In  Ref. \cite{Liang:2013sqa}, the authors studied the medium modifications of baryon properties in nuclear many-body system, especially in $\Lambda$ hypernuclei in the Friedberg-Lee model.  In  Ref. \cite{Jeong:2016qlk}, using the QCD sum-rule approach, the authors determined the self-energies and the energy of the quasi-$\Sigma$, -$\Lambda$ hyperons and quasi-neutron states and their density behaviour in neutron matter. The finite-density QCD sum-rule approach was also applied in  Ref. \cite{Jin:1994bh} for the investigation of $\Sigma$ hyperons self-energies propagating in nuclear matter. The author of Refs. \cite{Wang:2011hta,Wang:2012xk} studied the masses, vector self-energies and residues of the heavy $\Lambda_Q$ and doubly heavy $\Xi_{QQ}$ and $\Omega_{QQ}$ baryons in the nuclear matter using the QCD sum rule approach, as well. In Ref. \cite{Drukarev:2013kga}, the QCD sum rule was applied in the instanton medium to calculate the polarization operator of the nucleon current.  The in-medium spectroscopy of heavy spin-1/2 and light decuplet baryons were discussed in  Refs. \cite{Azizi:2016dmr,Azizi:2016hbr} where the authors were applied the in-medium QCD sum rules to calculate the masses, vector self-energies and residues of these baryons and determine their density-dependent behaviour. In a study about charmed baryon $\Lambda_c$ in nuclear matter \cite{Ohtani:2017wdc}, the authors studied density dependences of the mass and self-energies of $\Lambda_c$ in nuclear matter in the parity projected QCD sum rule. In a recent review article on heavy hadrons in nuclear medium \cite{Hosaka:2016ypm}, the current studies were reviewed with a summary of the basic theoretical concepts of QCD, namely chiral symmetry, heavy quark spin symmetry, and the effective Lagrangian approach.

In the present work, we investigate the heavy spin-3/2 $\Sigma_Q^{*}$, $\Xi_Q^{*}$ and  $\Omega_Q^{*}$ baryons using the in-medium QCD sum rule approach. We discuss the shifts in some spectroscopic parameters of these particles due to nuclear matter. We also study variations of those parameters with respect to the changes in the density of the cold nuclear medium.

This work is organized in the following form. In Sec. II, for the $\Sigma_Q^{*}$, $\Xi_Q^{*}$ and $\Omega_Q^{*}$  baryons, the in-medium sum rules for masses, residues and vector self-energies are obtained. In Sec. III, the numerical analysis for physical observables under consideration are performed. We move some lengthy expressions  to the Appendix.

\section{Spectroscopic properties of spin-3/2 $\Sigma_Q^{*}$, $\Xi_Q^{*}$ and $\Omega_Q^{*}$ baryons in cold nuclear matter}
The following in-medium two-point correlation function in momentum space is used to extract the in-medium hadronic parameters of the baryons under consideration in terms of the in-medium operators of quarks and gluons:
\begin{eqnarray}
\Pi_{\mu\nu} (p)= i \int d^4 x e^{i p \cdot x}  \langle  \psi_0| T [\zeta_{\mu} (x) \bar{\zeta}_{\nu}(0)] | \psi_0\rangle, \label{eq:corre1}
\end{eqnarray}
where $p$ is the four momentum of the heavy flavoured baryon with spin $3/2$ and $T$ is the time ordering operator. The $ |\psi_0\rangle$ in this equation is the nuclear matter ground state and it is featured by the nuclear matter rest frame nucleon density $\rho_N$ and the in-medium four-velocity $u_{\mu}$. The colorless interpolating field $\zeta_{\mu}$ composed of  quark fields with the same quantum numbers of the $\Sigma_Q^{*}, \Xi_Q^{*}$ and $\Omega_Q^{*}$ baryons is
\begin{eqnarray}\label{}
\zeta_{\mu}&=&\kappa \epsilon^{abc} \Big\{(q_{1}^{aT}C\gamma_\mu q_{2}^{b})Q^{c} + (q_{2}^{aT}C\gamma_\mu Q^{b})q_{1}^{c} \nonumber \\
&+&  (Q^{aT}C\gamma_\mu q_{1}^{b})q_{2}^{c} \Big\},
\end{eqnarray}
where $a, b$ and $c$ are color indices, $T$ denotes a transpose in Dirac space and $C$ is the charge-conjugation operator. The values of the  normalization constant $\kappa$ and the light quark contents of different members are collected in Table \ref{tab:QF}.
\begin{table}[tbp]
\begin{tabular}{|c|c|c|c|c|}\hline \hline  &$\kappa$ & $q_{1}$ & $q_{2}$   \\ \hline
$\Sigma_Q^{*}$ & $\sqrt{2/3}$ & u & d  \\
$\Xi_Q^{*}$ & $\sqrt{2/3}$ & s & u \\ 
$\Omega_Q^{*}$ & $\sqrt{1/3}$ & s & s  \\ \hline \hline \end{tabular}
\caption{The values of the normalization constant $\kappa$ and the quark flavors $q_1$ and $q_2$ for the heavy baryons with $J^P=3/2^+$.}
\label{tab:QF}
\end{table}

The aforementioned correlation function is calculated on hadronic and QCD sides. In QCD representation, the calculations are done with the help of the operator product expansion (OPE) in medium. Firstly, the hadronic side is obtained by inserting a complete set of baryonic state with the same quantum numbers as the interpolating current. Then, the integral over four-$x$ is performed. As a result, we get
\begin{eqnarray}\label{phepi}
\Pi_{\mu\nu}^{HAD}=&-&\frac{{\langle}\psi_0|\zeta_{\mu}(0)|B_Q(p^*,s){\rangle}
{\langle}B_Q(p^*, s)|\bar{\zeta}_{\nu}(0)|\psi_0{\rangle}}{p^{*2}-m_{B_Q}^{*2}} \notag \\
&+&...,
\end{eqnarray}
where  $p^*$ is in-medium four-momentum, $m_{B_Q}^{*}$ is the modified mass of the $|B_Q(p^*,s){\rangle}$ heavy  baryon state with spin $s$ in nuclear matter,  and dots show contributions of higher resonances and continuum states. The in-medium residue or coupling strength of the heavy baryon, $\lambda_{B_Q}^{*} $, can be defined through the matrix element  
\begin{eqnarray}\label{spinor}
{\langle}\psi_0|\zeta_{\mu}(0)|B_Q(p^*,s){\rangle} &=& \lambda_{B_Q}^{*} u_{\mu} (p^*,s),
\end{eqnarray}
where $u (p^*,s)$ is the Rarita-Schwinger spinor. The hadronic side of the correlation function is evolved by inserting Eq. (\ref{spinor}) into Eq. (\ref{phepi}) and summing over the spins of the $B_Q$ baryon state.

Before going further, it should be mentioned that the current $\zeta_{\mu}$ couples to both the spin-$1/2$ and the spin-$3/2$ states. But only the contributions of the spin-$3/2$ states must be considered and the  pollution of spin-$1/2$ states should be removed. To this end, the matrix element of $\zeta_{\mu}$ between the spin-$1/2$ and the in-medium ground state is parameterized as 
\begin{equation}\label{}
{\langle}\psi_0|\zeta_{\mu}(0)|\frac{1}{2} (p^*){\rangle} = \Big (K_1 p{*}_{\mu} + K_2 \gamma_{\mu} \Big ) u (p^*),
\end{equation}
where $K_1$ and $K_2$ are some constants. Using the condition $\zeta_{\mu} \gamma^{\mu}=0$, one can  immediately calculate the constant $K_1$ in terms of $K_2$,
\begin{equation}\label{SpinHalf}
{\langle}\psi_0|\zeta_{\mu}(0)|\frac{1}{2} (p^*){\rangle} = K_2 \Big (-\frac{4} {m_{1/2}^{*}}p^{*}_{\mu} +  \gamma_{\mu} \Big ) u (p^*),
\end{equation}
where $m_{1/2}^{*}$ is the in-medium mass of the spin-$1/2$ heavy baryonic state. As is seen from  Eq. (\ref{SpinHalf}), the spin-$1/2$ state's contributions are proportional to the  $p^{*}_{\mu} $ and $ \gamma_{\mu}$. Therefore, to eliminate these unwanted contributions, 
the Dirac matrices are ordered in a specific way and the contributions of the spin-1/2 particles are set to zero.

To proceed,  the Eq. (\ref{spinor}) is inserted into   Eq. (\ref{phepi}) and  summation over spins of the Rarita-Schwinger spinor is applied,  
\begin{eqnarray}\label{Rarita}
\sum_s  u_{\mu} (p^*,s)  \bar{u}_{\nu} (p^*,s) &= &-(\!\not\!{p^*} + m^{*}_{B_Q})\Big[g_{\mu\nu} -\frac{1}{3} \gamma_{\mu} \gamma_{\nu} \nonumber \\
&-& \frac{2p^*_{\mu}p^*_{\nu}}{3m^{*2}_{B_Q}} +\frac{p^*_{\mu}\gamma_{\nu}-p^*_{\nu}\gamma_{\mu}}{3m^{*}_{B_Q}} \Big].
\end{eqnarray}
As a result, we get
\begin{eqnarray}\label{}
\Pi_{\mu\nu}^{HAD}(p)&=&\frac{\lambda_{B_Q}^{*} \bar{\lambda}_{B_Q}^{*}(\!\not\!{p^*} + m^{*}_{B_Q})}{p^{*2}-m_{B_Q}^{*2}} \Big[g_{\mu\nu} -\frac{1}{3} \gamma_{\mu} \gamma_{\nu} \nonumber \\
&-& \frac{2p^*_{\mu}p^*_{\nu}}{3m^{*2}_{B_Q}} +\frac{p^*_{\mu}\gamma_{\nu}-p^*_{\nu}\gamma_{\mu}}{3m^{*}_{B_Q}} \Big]+ ...,
\end{eqnarray}
where $p_{\mu}^*=p_{\mu}-\Sigma_{\mu, v}$ and $m^{*}_{B_Q}=m_{B_Q}+\Sigma^S$. Here, $\Sigma_{\mu, \nu}$ and $\Sigma^S$ are the in-medium vector and scalar self energies, respectively. These physical observables can be calculated using  QCD sum rules. The two independent  vectors: the four-momentum of the particle, $p_{\mu}$, and the four-velocity of the medium, $u_{\mu}$, have contributions to the vector self-energy in the following way:
\begin{equation}\label{}
\Sigma_{\mu,v}=\Sigma_{v} u_{\mu} + \Sigma'_{v}p_{\mu},
\end{equation}
where $\Sigma_{v} $ and $ \Sigma'_{v}$ are constants and  due to its small contribution, $ \Sigma'_{v}$ can be ignored. 

Infinite nuclear matter has a natural rest frame that for the spatial components, the expectation value of the baryon current is zero but the time component of it in the same frame is the baryon density $\rho_N$. 
In this study, the calculations are operated in the rest frame of the nuclear medium, i.e. $u_{\mu}=(1,0)$. After some manipulations, the physical part of the correlation function can be decomposed in terms of different structures as
 \begin{eqnarray}\label{}
\Pi_{\mu\nu}^{HAD}(p_0, \vec{p})&=& \lambda_{B_Q}^{*} \bar{\lambda}_{B_Q}^{*}\frac{(\!\not\!{p}-\Sigma_{v}\!\not\!{u}+m^{*}_{B_Q})}{p^2+\Sigma_{v}^2 - 2p_0\Sigma_{v}-m^{*2}_{B_Q}}\Big[  g_{\mu\nu} \nonumber \\
&&-\frac{1}{3}\gamma_{\mu}\gamma_{\nu} -\frac{2}{3 m^{*2}_{B_Q}}\Big(p_{\mu}p_{\nu}-\Sigma_{v}p_{\mu}u_{\nu} \nonumber \\
&&-\Sigma_{v}u_{\mu}p_{\nu}+\Sigma^2_{v}u_{\mu}u_{\nu}\Big) +\frac{1}{3 m^{*}_{B_Q}}\Big(p_{\mu}\gamma_{\nu} \nonumber \\
&&-\Sigma_{v}u_{\mu}\gamma_{\nu}-p_{\nu}\gamma_{\mu}+\Sigma_{v}u_{\nu}\gamma_{\mu}\Big)\Big] + ....
\end{eqnarray}
After ordering of the Dirac matrices as $ \gamma_{\mu}\!\not\!{p}\!\not\!{u}\gamma_{\nu} $ and removing the pollution of the spin-1/2 states by setting the terms having  $ \gamma_{\mu}$ at the beginning and $ \gamma_{\nu}$ at the end and those that are proportional to $p_{\mu}$ and $p_{\nu}$ to zero, we have
\begin{eqnarray}\label{}
\Pi_{\mu\nu}^{HAD}(p_0, \vec{p})&=& 
\frac{\lambda_{B_Q}^{*} \bar{\lambda}^{*}_{B_Q}}{(p_0-E_p)(p_0-\bar{E}_p)}\Big[ m^{*}_{B_Q} g_{\mu\nu}+g_{\mu\nu}\!\not\!{p} \nonumber \\
&&-\Sigma_{v}g_{\mu\nu}\!\not\!{u} - \frac{4\Sigma^{2}_{v}}{3 m^{*}_{B_Q}} u_{\mu} u_{\nu}-\frac{2 \Sigma^{2}_{v}}{3 m^{*2}_{B_Q}} u_{\mu} u_{\nu}\!\not\!{p}\nonumber \\
&& + \frac{2\Sigma^{3}_{v}}{3 m^{*2}_{B_Q}} u_{\mu} u_{\nu}\!\not\!{u} \Big] + ...,
\end{eqnarray}
where $p_0=p\cdot u$ is the energy of the quasi-particle and $\Sigma_{v}$ is the vector self energy. In the mean field approximation, $E_p=\Sigma_{v}+\sqrt{p^2+ m^{*2}_{D}}$ is the positive energy pole and $\bar{E}_p=\Sigma_{v}-\sqrt{p^2+ m^{*2}_{D}}$ is the negative energy pole. 

In terms of the spectral densities the hadronic part reads
\begin{equation}\label{}
\Pi_{\mu\nu}^{HAD}(p_0, \vec{p})=\frac{1}{2\pi i} \int_{-\infty}^{\infty} d\omega \frac{\Delta \rho_{\mu\nu}^{HAD}(p_0, \vec{p})}{\omega-p_0},
\end{equation}
where the corresponding hadronic spectral density is found as
\begin{eqnarray}\label{}
\Delta \rho_{\mu\nu}^{HAD}(p_0, \vec{p})&=&\frac{2\pi i}{2\sqrt{m_{B_Q}^{*2}+\vec{p}^2}} \lambda_{D}^{*} \bar{\lambda}_{B_Q}^{*}\Big[ m^{*}_{B_Q} g_{\mu\nu}+g_{\mu\nu}\!\not\!{p} \nonumber \\
&&-\Sigma_{v}g_{\mu\nu}\!\not\!{u} - \frac{4 \Sigma^{2}_{v}}{3 m^{*}_{B_Q}} u_{\mu} u_{\nu}-\frac{2 \Sigma^{2}_{v}}{3 m^{*2}_{B_Q}} u_{\mu} u_{\nu}\!\not\!{p}\nonumber \\
&& + \frac{2 \Sigma^{3}_{v}}{3 m^{*2}_{B_Q}} u_{\mu} u_{\nu}\!\not\!{u} \Big] \Big[ \delta(\omega - E_p) - \delta(\omega- \bar{E}_p) \Big] . \nonumber \\
\end{eqnarray}
The correlation function can be obtained multiplying the above spectral density with the weight function $(\omega-\bar{E}_p)e^{\frac{-\omega^2}{M^2}}$  and performing the integral 
\begin{equation}\label{integ}
\Pi_{\mu\nu}^{HAD}(p_0, \vec{p})= \int_{-\omega_0}^{\omega_0} d\omega \Delta \rho_{\mu\nu}^{HAD}(\omega, \vec{p})(\omega-\bar{E}_p)e^{-\frac{\omega^2}{M^2}},
\end{equation}
to exclude the negative-energy pole contribution. Here $\omega_0$ is the threshold parameter and $M^2$ is the Borel mass parameter to be fixed. After performing the integral in Eq. (\ref{integ}), we get the Hadronic side of the correlation function in terms of the related structures as
\begin{eqnarray}
\Pi_{\mu\nu}^{HAD}(p_0, \vec{p})&=& \lambda_{B_Q}^{*} \bar{\lambda}_{B_Q}^{*} e^{-E^2_p/M^2}\Big[ m^{*}_{B_Q} g_{\mu\nu}+g_{\mu\nu}\!\not\!{p}  \nonumber \\
&-& \Sigma_{v}g_{\mu\nu}\!\not\!{u}-\frac{4 \Sigma^{2}_{v}}{3 m^{*}_{B_Q}} u_{\mu} u_{\nu}-\frac{2 \Sigma^{2}_{v}}{3 m^{*2}_{B_Q}} u_{\mu} u_{\nu}\!\not\!{p} \nonumber \\
&+& \frac{2 \Sigma^{3}_{v}}{3 m^{*2}_{B_Q}} u_{\mu} u_{\nu}\!\not\!{u} \Big]. \nonumber \\
\end{eqnarray}

The other side of the sum rule  is the direct calculation of the correlation function using the interpolating currents via the OPE. In a space-like region, the short and long range effects are separated and are written in terms of different operators having various mass dimensions in nuclear medium   \cite{Cohen:1994wm},
\begin{equation}
\label{ }
\Pi(p^2)=\Sigma_n C^{i}_n (p^2) \langle \hat{O}_n\rangle_{\rho_N},
\end{equation} 
where the $C^{i}_n$ are corresponding Wilson coefficients and the $ \langle \hat{O}_n\rangle_{\rho_N} =  \langle  \psi_0|\hat{O}_n | \psi_0\rangle$
denote the in-medium expectation values of QCD operators, i.e. the condensates. 

The correlation function on QCD side can also be decomposed in terms of the involved structures as
\begin{eqnarray}
\Pi_{\mu\nu}^{QCD} (p_0, \vec{p})& = & \Pi_1 g_{\mu\nu}+ \Pi_2 g_{\mu\nu}\!\not\!{p} + \Pi_3 g_{\mu\nu}\!\not\!{u}  \nonumber \\
 & + &  \Pi_4 u_{\mu} u_{\nu}+ \Pi_5 u_{\mu} u_{\nu}\!\not\!{p} +  \Pi_6 u_{\mu} u_{\nu}\!\not\!{u}, \nonumber \\
\end{eqnarray}
where the coefficients of the above structures, $\Pi_i$ with $i=1, ..., 6$, are scalar functions of the invariants $p^2$ and $p\cdot u$. In the vacuum limit only the $ g_{\mu\nu}$ and $g_{\mu\nu}\!\not\!{p}$ structures remain non-vanishing and the  coefficients $ \Pi_1$ and  $\Pi_2$ become functions of $p^2$. In the nuclear medium, the $\Pi_{i}$ functions on QCD side can also be written as
\begin{equation}
\label{ }
\Pi_i=\frac{1}{2\pi i} \int_{-\infty}^{-\infty} d\omega \frac{\Delta \rho_i}{\omega-p_0},
\end{equation}
where the spectral densities  $\Delta \rho_i $ are the imaginary parts of $\Pi_i$ functions. 

The main aim in the following is to calculate these spectral densities. The first step is to insert the explicit expressions of the interpolating currents and contracting out all quark pairs using the Wick's theorem. This leads to a result in terms of the heavy and light quark propagators in $x$-space. We move the expressions obtained in this step to the Appendix. By using the $x$-expressions of the
 heavy and light propagators and applying the well-known formula,
\begin{eqnarray}
\label{ }
\frac{1}{(x^2)^m}&=&\int \frac{d^Dk }{(2\pi)^D}e^{-ik \cdot x}i(-1)^{m+1}2^{D-2m}\pi^{D/2} \nonumber \\
&\times& \frac{\Gamma[D/2-m]}{\Gamma[m]}\Big(-\frac{1}{k^2}\Big)^{D/2-m},
\end{eqnarray}
we obtain an expression with three four-dimensional integrals. We can easily perform integral over four$-x$ leading to a Dirac Delta. The resultant Dirac Delta function helps us to perform the second four-integral. The final four-integral is performed using the Feynman parametrization method. This leads to the integrals of the form
\begin{equation}
\label{ }
\int d^4 \ell\frac{(\ell^2)^m}{(\ell^2+\Delta)^n}=\frac{i\pi^2 (-1)^{m-n} \Gamma[m+2]\Gamma[n-m-2]}{\Gamma[2]\Gamma[n] (-\Delta)^{n-m-2}}.
\end{equation} 
To extract the imaginary parts corresponding to different structures, the following relation is used:
\begin{equation}
\label{ }
\Gamma\Big[\frac{D}{2}-n\Big]\Big(-\frac{1}{\Delta}\Big)^{D/2-n}=\frac{(-1)^{n-1}}{(n-2)!}(-\Delta)^{n-2}ln[-\Delta].
\end{equation} 
where the condition $\Delta>0$ brings constraints on the limits of the integral over the Feynman parameter and $ln[-\Delta]=i\pi+ln[\Delta]$.

To proceed, we multiply the obtained spectral densities by the same weight function as the hadronic side and apply the integral over $w$ from $-w_0$ to $+w_0$. Then we put  $w_0=\sqrt{s_0^*}$, with $s_0^*$ being the in-medium continuum threshold. After some variable changing, the correlation functions $\Pi_{i}^{QCD}(s_0^{*},M^2)$ in the Borel scheme are obtained,
\begin{equation}
\label{ }
\Pi_{i}^{QCD}(s_0^{*},M^2)=\int_{m_Q^2}^{s_0^{*}} ds \rho_i^{QCD} (s) e^{\frac{-s}{M^2}}.
\end{equation}
As examples , we present the explicit forms of the above spectral densities corresponding to the structure $g_{\mu\nu}$ for $\Sigma^{*}_b$, $\Xi^{*}_b$ and $\Omega^{*}_b$ baryons below:
\begin{widetext}
\begin{eqnarray}
\label{ RhoSigma}
\rho^{\Sigma^{*}_b}_{1}(s)&=& \frac{1}{576  \pi^4} \int^1_0  \frac{dz}{(z-1)^2}\Bigg\{ 84 m_b z^2 (z-3)  [m_b^2+s (z-1)]^2 - 384 \pi^2 z (z-1)^2  \Big[m_q m_b+3 \big[m_b^2+2 s (z-1)\big]\Big]   \nonumber \\
&\times& \Big( {\langle} \bar{u} u{\rangle}_{\rho_{N}} + {\langle} \bar{u}u{\rangle}_{\rho_{N}} \Big) +2304 \pi^2 m_q \sqrt{s_0^*} z (z-1)^3  \Big( {\langle}u^{\dag}u{\rangle}_{\rho_N}  +  {\langle}d^{\dag}d{\rangle}_{\rho_N}\Big) + 3 \pi^2  (z-1)^2 [256 z^2-256 z-3] \nonumber \\
&\times& \Big( {\langle} \bar{u}g_s\sigma Gu{\rangle}_{\rho_N} +  {\langle} \bar{d}g_s\sigma Gd{\rangle}_{\rho_N} \Big)  + 12 i \pi^2 (z-1)^2 (3 m_q-64 m_b z) \Big(  {\langle}u^{\dag}iD_0u{\rangle}_{\rho_N} + {\langle}d^{\dag}iD_0d{\rangle}_{\rho_N}\Big)  \nonumber\\
&+& \pi^2  m_b (3 z^4-z^3-13 z^2+5 z+2) \langle \frac{\alpha_s}{\pi} G^{2}\rangle_{\rho_N} + 12 \pi^2  (z-1)^2 (64 z^2 - 64 z + 3) \Big( {\langle}\bar{u}iD_0iD_0u{\rangle}_{\rho_{N}} \nonumber \\
&+& {\langle}\bar{d}iD_0iD_0d{\rangle}_{\rho_{N}}\Big) \Bigg\} \Theta[L(s,z)],
\end{eqnarray}
\begin{eqnarray}
\label{ RhoXi}
\rho^{\Xi^{*}_b}_{1}(s)&=& \frac{1}{576  \pi^4} \int^1_0  \frac{dz}{(z-1)^2}\Bigg\{- 84 m_b z^2 (z-3)  [m_b^2+s (z-1)]^2 + 48 \pi^2 z (z-1)^2 \Big[m_q m_b+3 \big[m_b^2+2 s (z-1)\big]\Big]   \nonumber \\
&\times& \Big( {\langle} \bar{u} u{\rangle}_{\rho_{N}} + {\langle} \bar{s}s{\rangle}_{\rho_{N}} \Big)  -288 \pi^2 m_q \sqrt{s_0^*} z (z-1)^3  \Big( {\langle}u^{\dag}u{\rangle}_{\rho_N}  +  {\langle}s^{\dag}s{\rangle}_{\rho_N}\Big) - 3 \pi^2  (z-1)^2 [32 z^2-32z-3] \nonumber \\
&\times& \Big( {\langle} \bar{u}g_s\sigma Gu{\rangle}_{\rho_N} +  {\langle} \bar{s}g_s\sigma Gs{\rangle}_{\rho_N} \Big)  + 12 i \pi^2 (z-1)^2 (-3m_q+8m_bz)\Big({\langle}u^{\dag}iD_0u{\rangle}_{\rho_N} + {\langle}s^{\dag}iD_0s{\rangle}_{\rho_N}\Big) \nonumber\\
&+& \pi^2  m_b (-3 z^4+z^3+13 z^2-5 z-2) \langle \frac{\alpha_s}{\pi} G^{2}\rangle_{\rho_N} - 12 \pi^2  (z-1)^2 (8z^2 - 8 z + 3) \Big({\langle}\bar{u}iD_0iD_0u{\rangle}_{\rho_{N}}  \nonumber \\
&& + {\langle}\bar{s}iD_0iD_0s{\rangle}_{\rho_{N}} \Big) \Bigg\} \Theta[L(s,z)],
\end{eqnarray}
\begin{eqnarray}
\label{ RhoOmega}
\rho^{\Omega^{*}_b}_{1}(s)&=& \frac{1}{576  \pi^4} \int^1_0  \frac{dz}{(z-1)^2}\Bigg\{ -96 m_b z^2 (z-3)  [m_b^2+s (z-1)]^2 + 768 \pi^2 z (z-1)^2  \big[m_q m_b + 3 [m_b^2 + 2 s (z-1])\big]  {\langle} \bar{s} s{\rangle}_{\rho_{N}} \nonumber \\
&-& 4608  \pi^2 m_q \sqrt{s_0^*} z ( z-1)^3 {\langle}s^{\dag}s{\rangle}_{\rho_N} -6 \pi^2 (z-1)^2 (-24 - 235 z + 256 z^2)  {\langle} \bar{s}g_s\sigma Gs{\rangle}_{\rho_N} + 24 i \pi^2 (z-1)^2 [64 m_b z  \nonumber \\
&+& 3 m_q (-8 + 7 z)] {\langle}s^{\dag}iD_0s{\rangle}_{\rho_N} +  \pi^2 m_b (-16 - 12 z + 69 z^2 + 15 z^3 - 24 z^4)  \langle \frac{\alpha_s}{\pi} G^{2}\rangle_{\rho_N} -24 \pi^2 (-1 + z)^2 (24 - 85 z \nonumber \\
&+& 64 z^2) {\langle}\bar{s}iD_0iD_0s{\rangle}_{\rho_{N}} \Bigg\} \Theta[L(s,z)],
\end{eqnarray}
\end{widetext}
where $\Theta[L(s,z)]=[-m_b^2 z + s (z - z^2)]$.

As the final step the coefficients of different structures from the hadronic and QCD sides of the correlation function are matched. As a result, we find the following sum rules that will be used in numerical calculations of the mass, residue and vector self-energy of the baryons under consideration:
\begin{eqnarray}\label{SumR}
m^{*}_{B_Q}\lambda^{*2}_{B_Q} e^{-\frac{E_p^2}{M^2}}& = & \Pi^{QCD}_{1}, \nonumber  \\
\lambda^{*2}_{B_Q} e^{-\frac{E_p^2}{M^2}}& = & \Pi^{QCD}_{2}, \nonumber  \\
-\Sigma_{\upsilon}\lambda^{*2}_{B_Q} e^{-\frac{E_p^2}{M^2}}& = & \Pi^{QCD}_{3}.
\end{eqnarray}

 \section{Numerical analyses}
 For the numerical analyses of the obtained sum rules in the previous section, we need the numerical values of quark and baryon masses, saturated nuclear matter density as well as the quark, gluon and mixed condensates. They are presented in Table II. 
 \begin{widetext}
 
\begin{table}[ht!]
\centering
\begin{tabular}{ |c|c|c|}
\hline \hline
Parameter  &  Numeric & Unit  \\ \hline
$ m_{\Sigma^{*}_{b}} $; $ m_{\Sigma^{*}_{c}} $   &  $ 5.832  $; $ 2.518  $   & $GeV$    \\   
$ m_{\Xi^{*}_{b}} $; $ m_{\Xi^{*}_{c}} $   &  $ 5.953 $ ; $ 2.646  $   & $GeV$    \\  
$ m_{\Omega^{*}_{b}} $; $ m_{\Omega^{*}_{c}} $   &  $ 6.080   $;  $ 2.766 $   & $GeV$    \\ 
$ m_q$; $ \sigma_N$ &  $ 0.00345  $;  $ 0.059  $  & $GeV$    \\  
$  m_0^2 $ & $ 0.8 $ & $GeV^2$    \\
$ \rho_{N}^{sat} $   &  $0.11^3 $  & $GeV^3$    \\  
$\langle q^{\dag}q\rangle_{\rho_N} $; $\langle s^{\dag}s\rangle_{\rho_N} $ &  $\frac{3}{2}\rho_{N}$; 0  & $GeV^3$    \\  
$ \langle\bar{q}q\rangle_{0} $; $ \langle\bar{s}s\rangle_{0} $ &  $ (-0.241)^3$; $ 0.8 \langle\bar{q}q\rangle_{0} $& $GeV^3$    \\  
$ \langle\bar{q}q\rangle_{\rho_N} $; $ \langle\bar{s}s\rangle_{\rho_N} $&  $\langle\bar{q}q\rangle_{0}+ \frac{\sigma_N}{2 m_q} \rho_N  $; $\langle\bar{s}s\rangle_{0}+ y\frac{\sigma_N}{2 m_q} \rho_N  $ & $GeV^3$    \\  
$ \langle q^{\dag}iD_0q\rangle_{\rho_N}$; $ \langle s^{\dag}iD_0s\rangle_{\rho_N}$&  $0.18 \rho_N $; $ \frac{m_s \langle\bar{s}s\rangle_{\rho_N}}{4}+0.02 ~GeV \rho_N$& $GeV^4$   \\   
$ \Big\langle\frac{\alpha_s}{\pi}G^2\Big\rangle_{0}$; $\Big\langle \frac{\alpha_s}{\pi}G^2\Big\rangle_{\rho_N}$ & $ (0.33 \pm 0.04)^4 $; $\Big\langle \frac{\alpha_s}{\pi}G^2\Big\rangle_{0}-0.65~GeV \rho_N $   & $GeV^4$    \\   
$ \langle\bar{q}g_s\sigma Gq \rangle_{0}$; $ \langle\bar{q}g_s\sigma Gq \rangle_{\rho_N}$ & $ m_0^2 \langle\bar{q}q\rangle_{0}  $; $ \langle\bar{q}g_s\sigma Gq \rangle_{0} + 3 GeV^2 \rho_N$  & $GeV^5$    \\ 
$ \langle\bar{q}iD_0 iD_0q\rangle_{\rho_N} $; $  \langle s^{\dag}iD_{0}iD_{0}s\rangle_{\rho_N}$ & $ 0.3~GeV^2\rho_{N}-\frac{1}{8}\langle\bar{q}g_{s}\sigma Gq\rangle_{\rho_N}$;  $ 0.3 y ~GeV^2\rho_{N}-\frac{1}{8}\langle\bar{s}g_{s}\sigma Gs\rangle_{\rho_N}$& $GeV^5$    \\  
$y$ & $0.04\pm0.02$ ; $0.066\pm0.011\pm0.002$ & - \\
\hline \hline
\end{tabular}
\caption{Inputs parameters used in calculations \cite{PDG,Cohen:1991nk,Cohen:1994wm,Belyaev:1982cd,Ioffe:2005ym,Thomas:2012tg,Dinter:2011za,Alarcon:2011zs}.}
\end{table}

\end{widetext}

Additionally, it is necessary to fix two  auxiliary parameters:  the continuum threshold $s_0^*$ and Borel mass parameter $M^{2}$. The requirements for this aim  according to the standard prescriptions are: convergence of OPE, exceeding of the perturbative part over the non-perturbative one as well as the pole dominance in medium. We have an additional restriction on the continuum threshold parameter. It depends on the energy of the first excited state with the same quantum numbers. It means that, for the channels under consideration, while $m_{B_Q}$ is the ground state energy, the energy for the first excited state of the particle is $\sqrt{s^{*}_0}-m_{B_Q}$. These conditions lead to the intervals: $s^{*}_0= (m_{B_Q}+[0.3\div 0.5])^2$ GeV$^2$ for both $b-$ and $c-$ baryons, as well as, $M^{2}=[5\div 8]$ GeV$^2$ for $b-$baryons  and $M^{2}=[3\div 6]$ GeV$^2$ for $c-$baryons. To check the convergence, as examples, we show the contributions of the perturbative and different non-perturbative operators for the structure $g_{\mu\nu}$ and the $b-$ and $c-$quark cases  with respect to $M^2$ at fixed values of the continuum threshold and nuclear matter saturation density in figure 1.  From this figure, we see that the OPE nicely converges for both the $b$ and $c$ cases in $\Sigma_Q^{*}$ channel. Indeed, the perturbative part constitutes roughly $70\%$ and $80\%$ of the total contribution for $b$ and $c$ cases, respectively. Similar results are obtained in $\Xi_Q^{*}$ and $\Omega_Q^{*}$ channels, as well. In figure 2, as an example, we also show the pole contributions for both the $b$ and $c$ cases in $\Sigma_Q^{*}$ channel which amount $ (71-84)$ $\%$  and $ (50-72)$ $\%$, respectively satisfying another condition of the method.
\begin{widetext}

\begin{figure}[h!]
\label{fig1}
\centering
\begin{tabular}{cc}
\epsfig{file=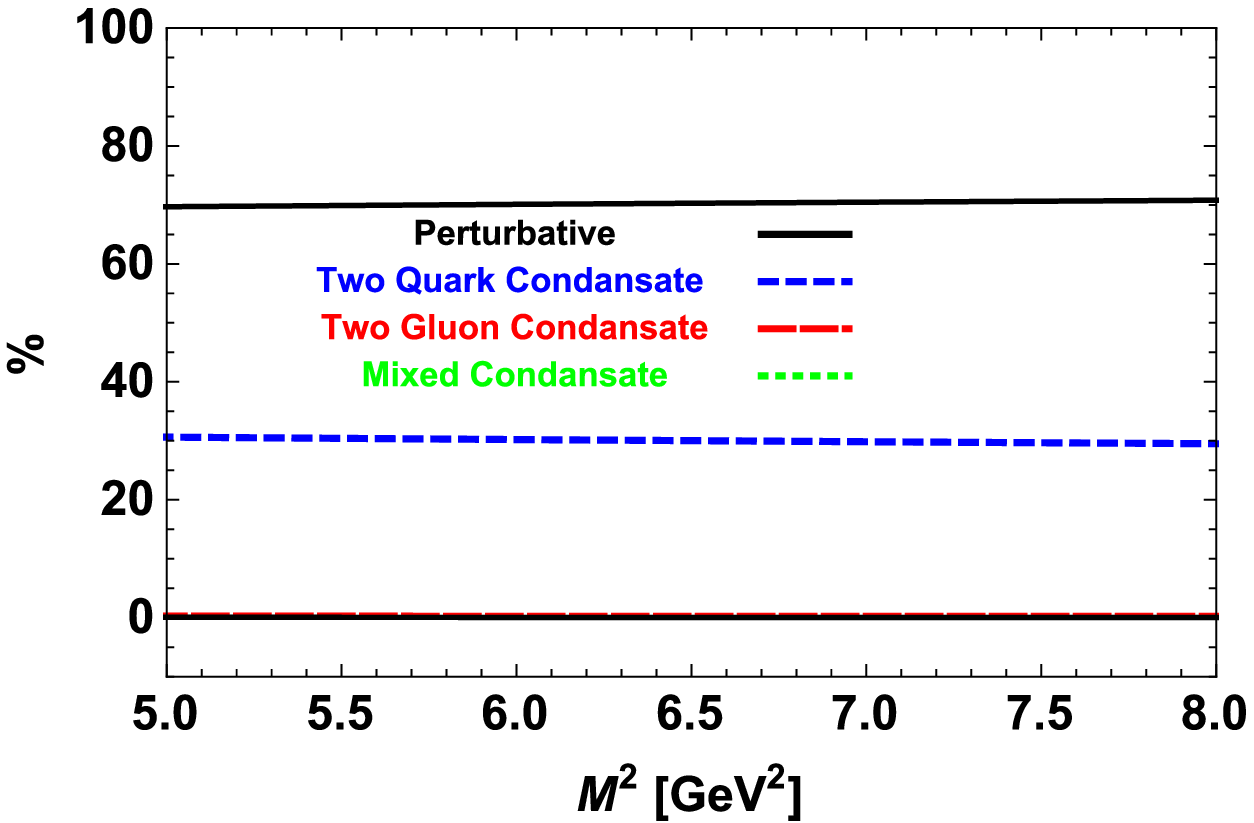,width=0.45\linewidth,clip=} &
\epsfig{file=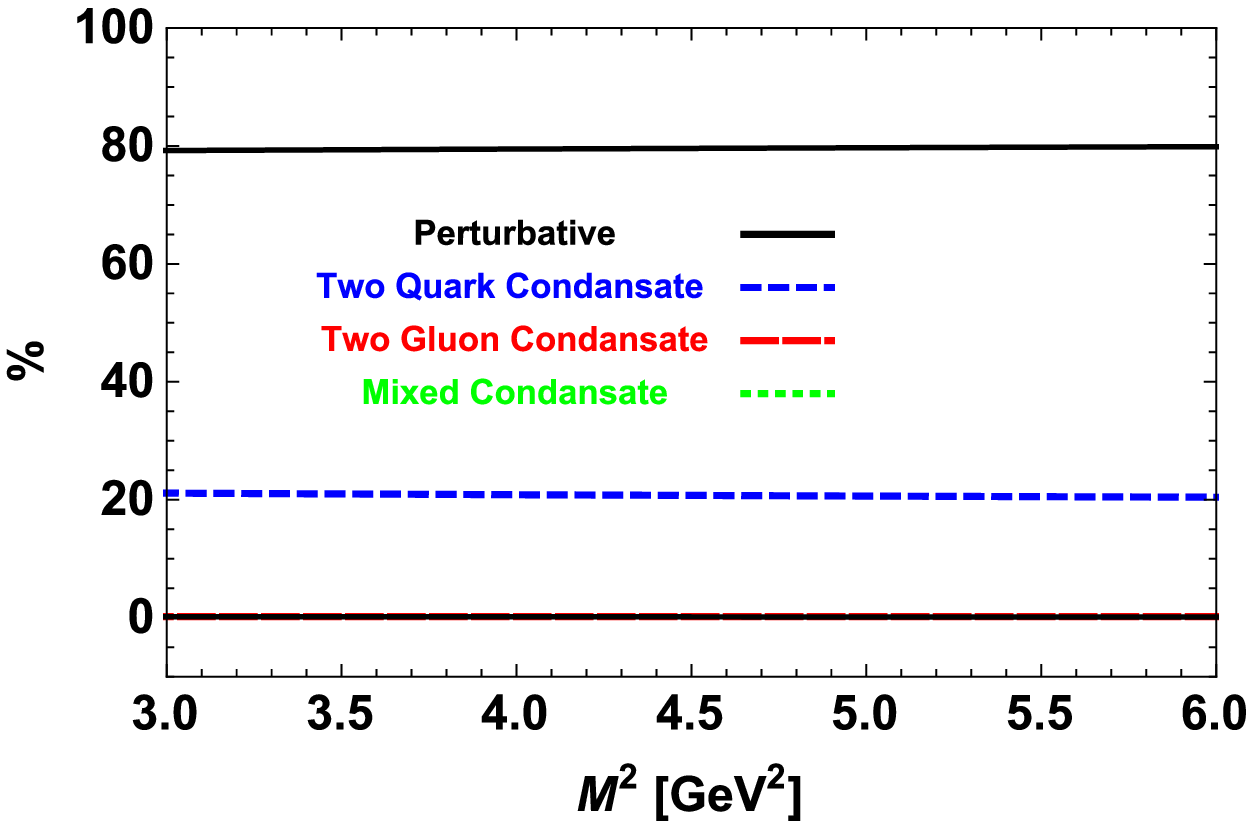,width=0.45\linewidth,clip=}  
\end{tabular}
\caption{Contributions of the perturbative, two-quark condensate, two-gluon condensate and mixed condensate in OPE for $b$-quark (left) and $c$-quark (right) in $\Sigma_Q^{*}$ channel in terms of $M^2$ at saturation density and different fixed values of in-medium continuum threshold.}
\end{figure}
\end{widetext}
\begin{widetext}

\begin{figure}[h!]
\label{}
\centering
\begin{tabular}{cc}
\epsfig{file=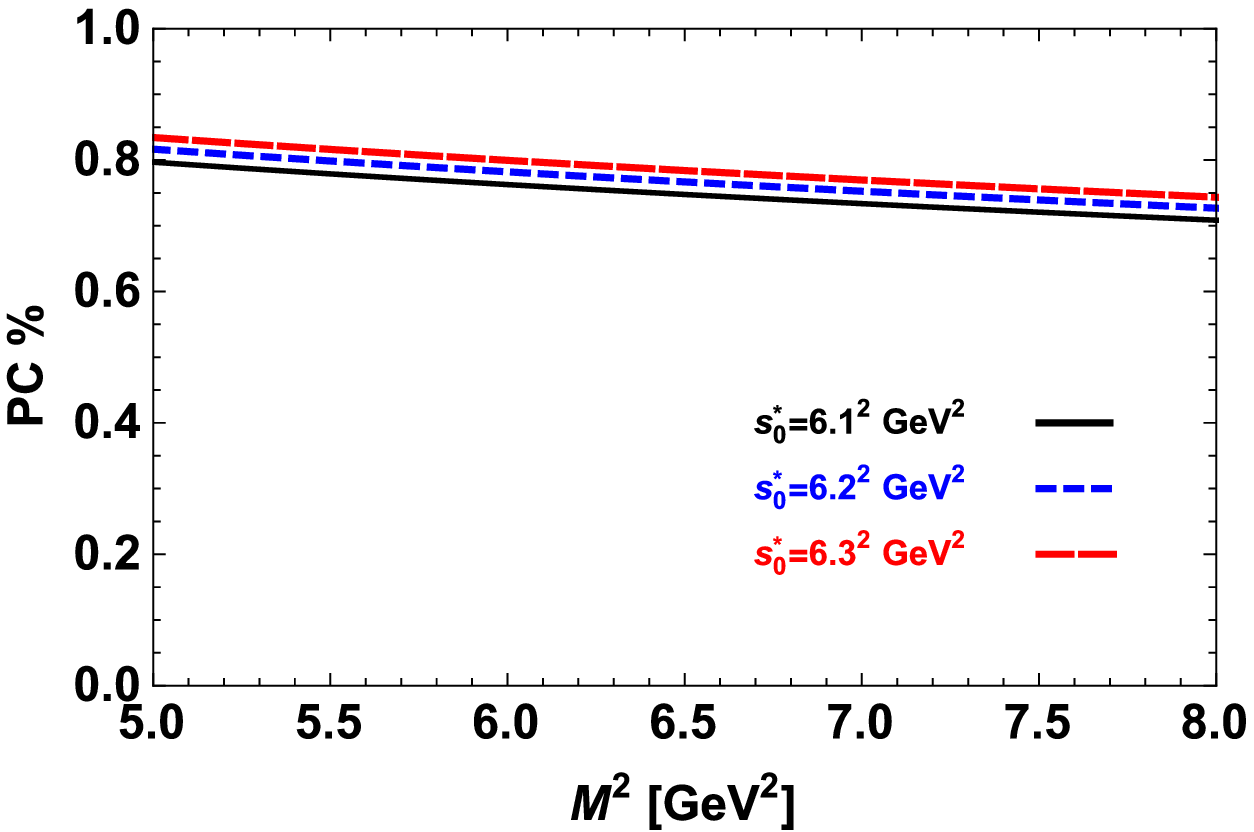,width=0.45\linewidth,clip=} &
\epsfig{file=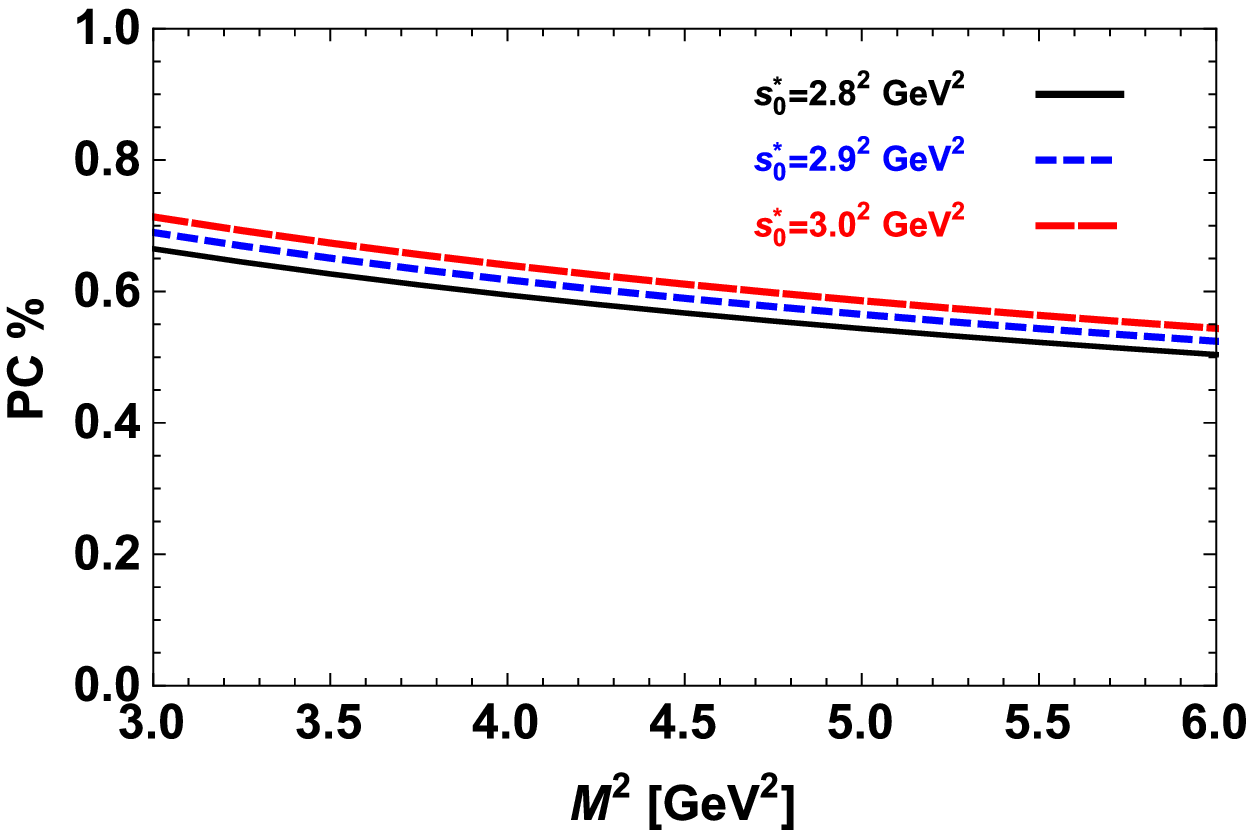,width=0.45\linewidth,clip=}  
\end{tabular}
\caption{The pole contributions in the $ \Sigma^{*}_{Q}$ channel for $b$-quark (left) and $c$-quark (right) as a function of  $M^2$ at saturation density and different fixed values of in-medium continuum threshold.}
\end{figure}
\end{widetext}

To show the density dependence of the in-medium energy ($E_p$),  mass ($ m^*_{\Sigma^{*}_{Q}}$) and vector self-energy ($\Sigma_{v}$) of $\Sigma_Q^{*}$ baryon, we plot the ratio of these quantities to the related vacuum mass ($ m_{\Sigma^{*}_{Q}}$) as a function of $\rho_N/\rho^{\textrm{sat}}_N$ at mean values of the Borel parameter and continuum threshold for $\Sigma_b^{*}$ baryon (left panel) and $\Sigma_c^{*}$ baryon (right panel) in figure 3.  In both channels, the  $E_p/ m_{\Sigma^{*}_{Q}}$ ratio is almost density independent.  On the other hand, as a result of sum rules in Eqs. (\ref{SumR}),  there is a coupled relation between the in-medium mass and the vector self energy of the considered baryon, while  $ m^*_{\Sigma^{*}_{Q}}/m_{\Sigma^{*}_{Q}}$ ratio is decreasing with increasing medium density, $\Sigma_{v}/m_{\Sigma^{*}_{Q}}$ ratio is increasing. The vector self energy is zero at $ \rho_N \rightarrow 0 $ limit as expected. The mass shift due to nuclear matter in $\Sigma_b^{*}$ channel is greater than  $\Sigma_c^{*}$ channel. In figure 4, we plot the Borel mass dependencies of the same ratios of physical quantities. As required, all quantities show good stability with respect to the changes of the Borel mass in both the $\Sigma_b^{*}$ (left panel) and $\Sigma_c^{*}$ channels (right panel). We observe similar behaviors but with relatively less amount of shifts due to the cold nuclear medium in other  baryonic channels.
 
\begin{widetext}

\begin{figure}[h!]
\label{fig2}
\centering
\begin{tabular}{cc}
\epsfig{file=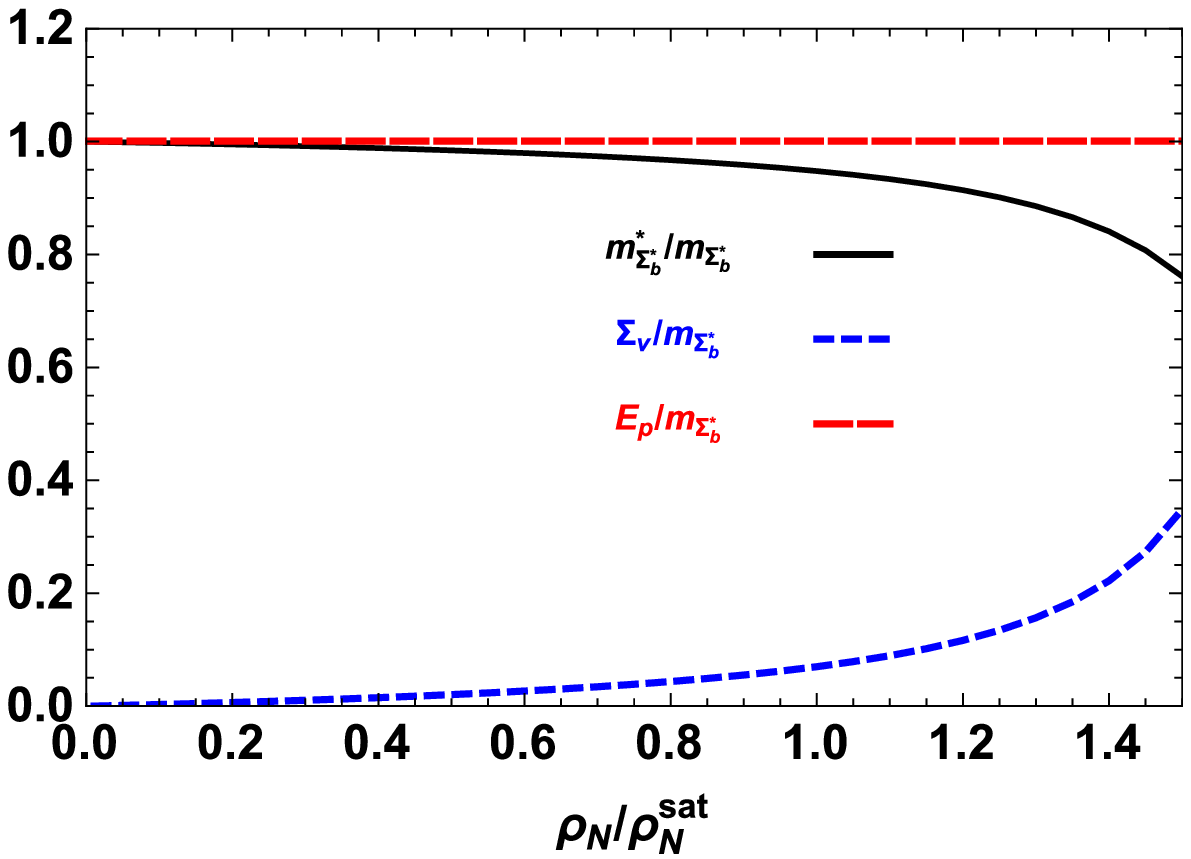,width=0.45\linewidth,clip=} &
\epsfig{file=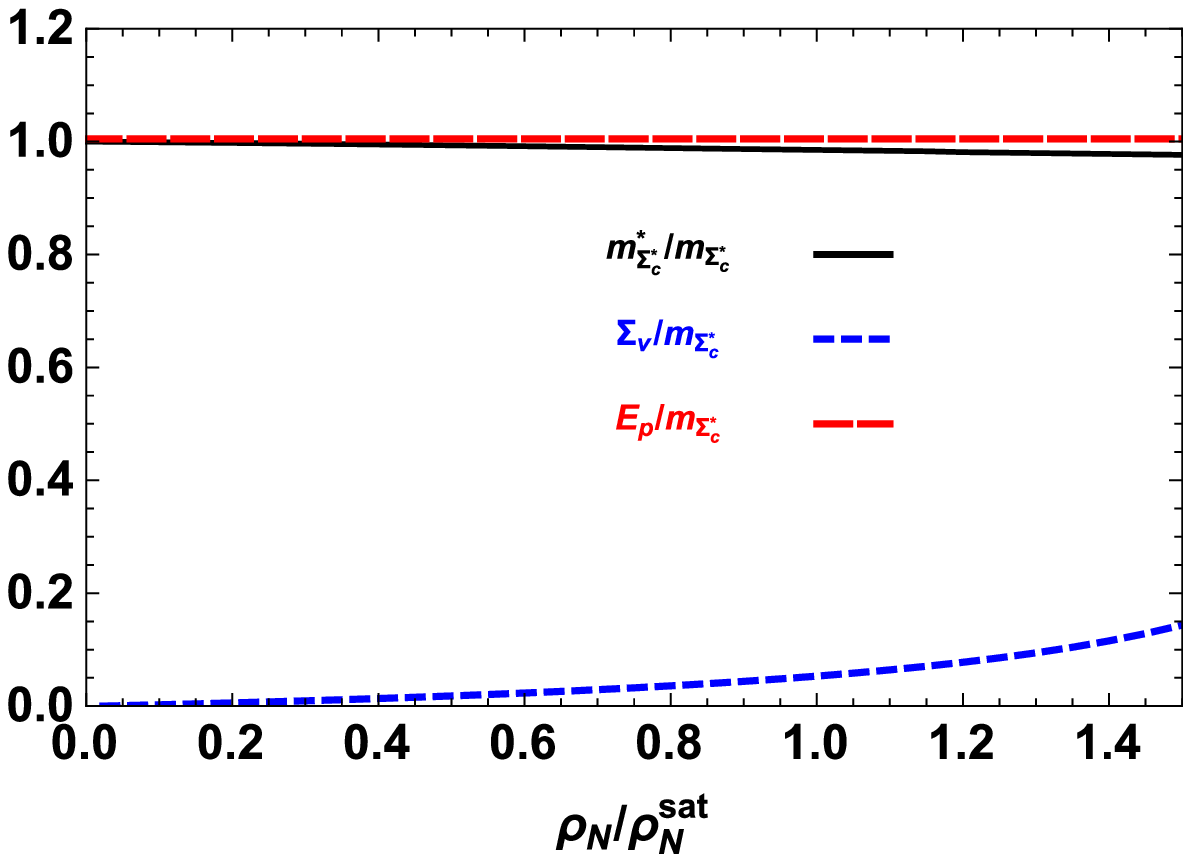,width=0.45\linewidth,clip=}  
\end{tabular}
\caption{The ratios  $ m^*_{\Sigma^{*}_{Q}}/m_{\Sigma^{*}_{Q}}$, $\Sigma_{v}/m_{\Sigma^{*}_{Q}}$ and $E_p /m_{\Sigma^{*}_{Q}}$ in terms of  $\rho_N/\rho^{\textrm{sat}}_N$ at average values of the continuum threshold and Borel parameter for $b-$quark (left panel) and $c-$quark (right panel).}
\end{figure}
\begin{figure}[h!]
\label{fig3}
\centering
\begin{tabular}{cc}
\epsfig{file=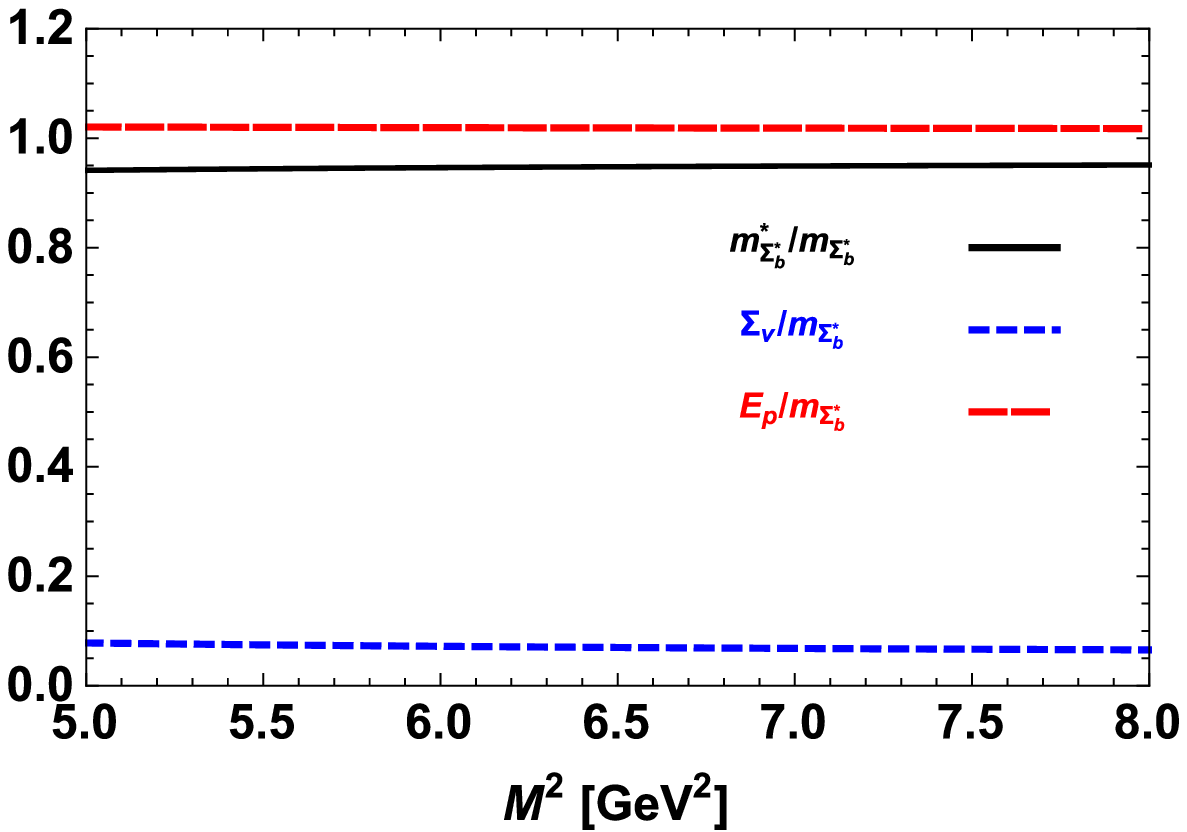,width=0.45\linewidth,clip=} &
\epsfig{file=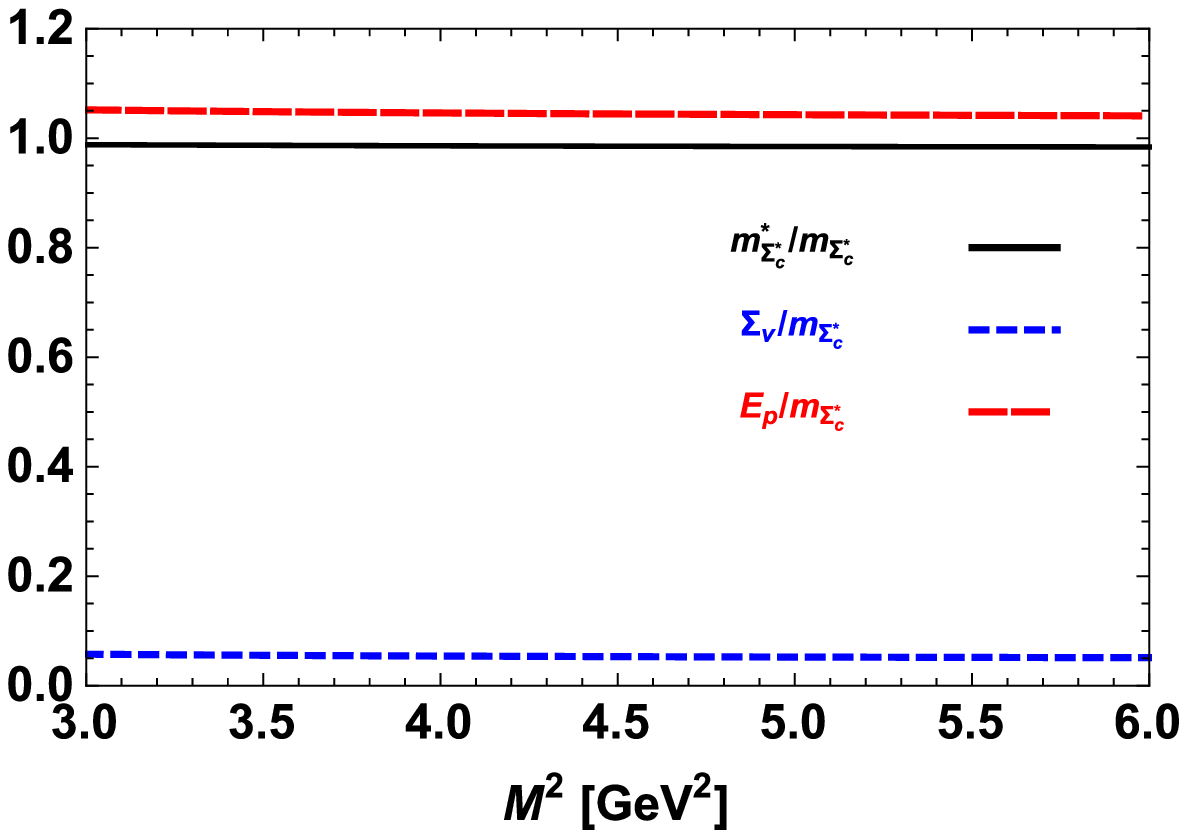,width=0.45\linewidth,clip=}  
\end{tabular}
\caption{Same ratios with figure 3 but as a function of $M^2$ at the saturated nuclear matter density.}
\end{figure}
\end{widetext}

\begin{widetext}

\begin{figure}[h!]
\label{fig4}
\centering
\begin{tabular}{cc}
\epsfig{file=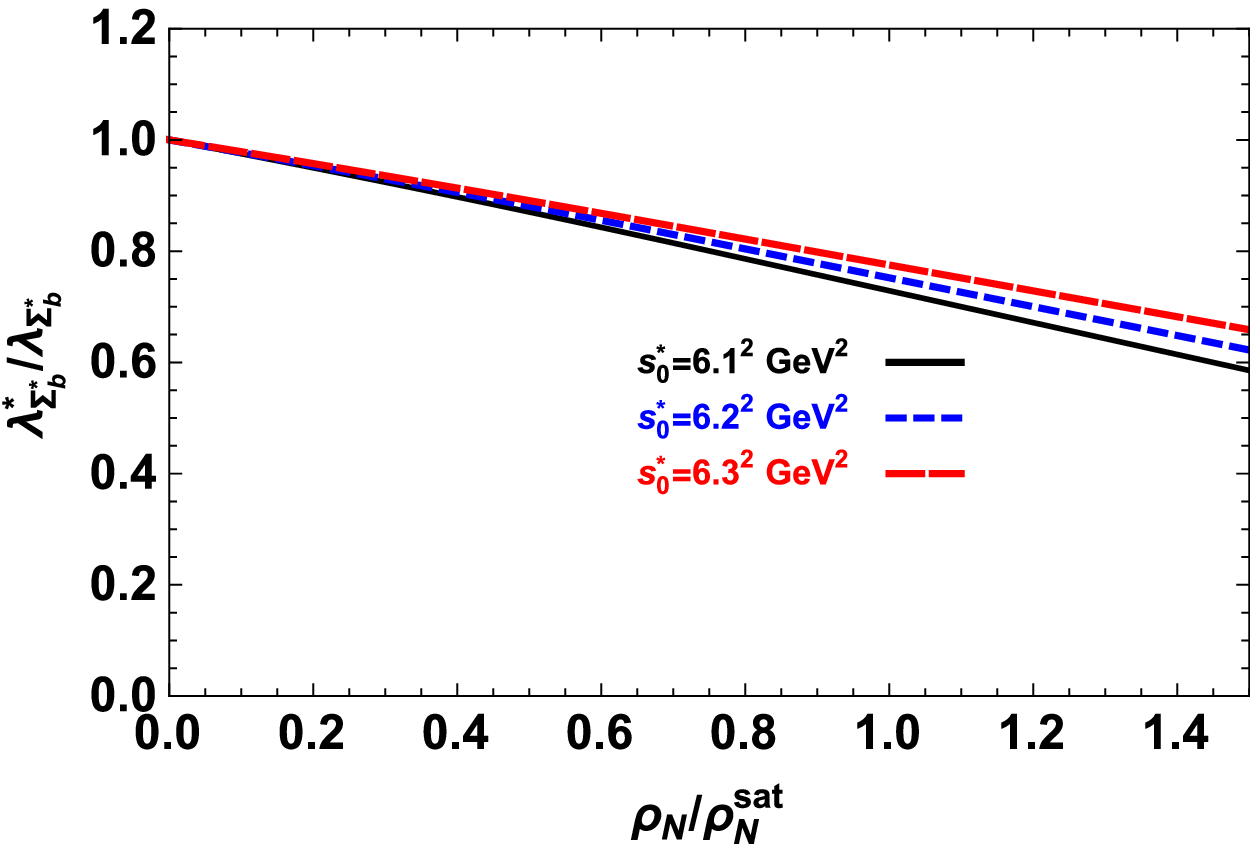,width=0.45\linewidth,clip=} &
\epsfig{file=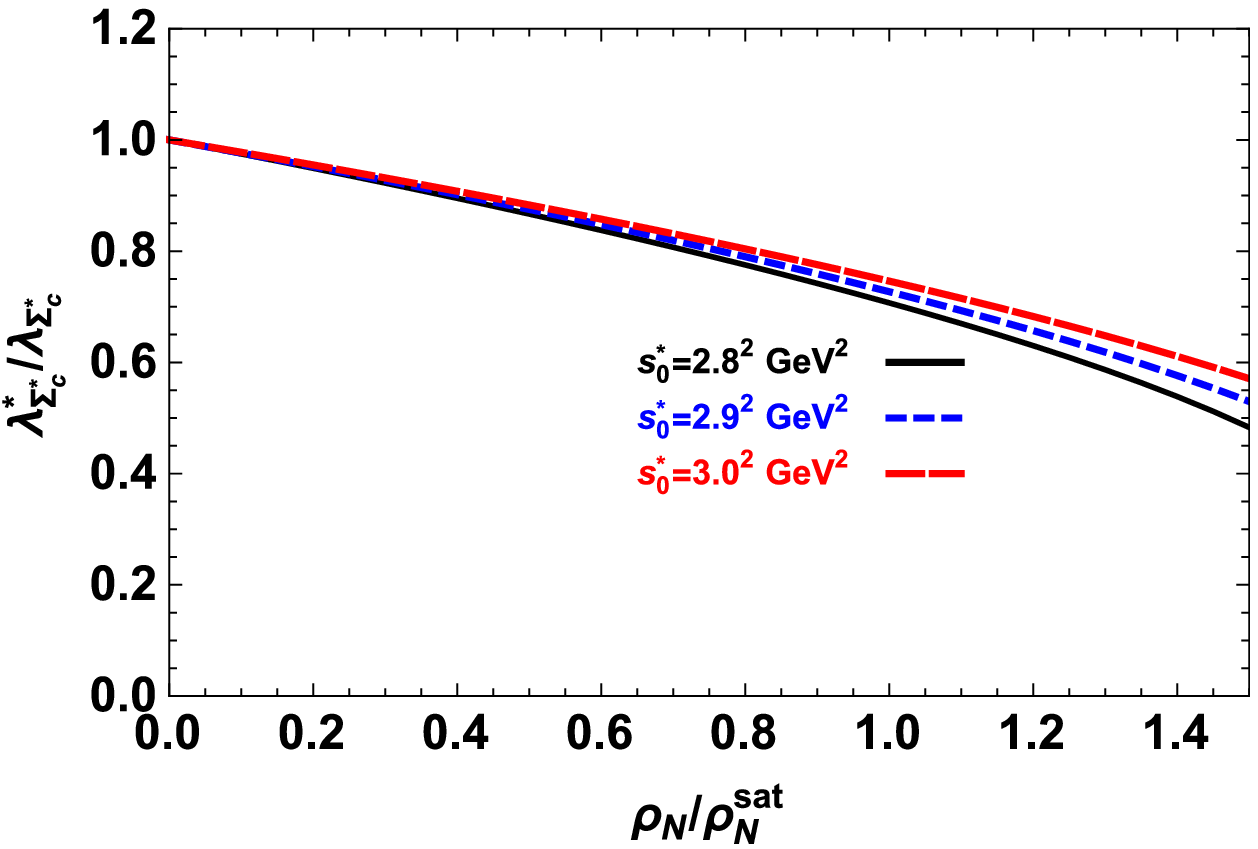,width=0.45\linewidth,clip=}  
\end{tabular}
\caption{The ratios  $\lambda^{*}_{\Sigma_{b}^{*}}/\lambda_{\Sigma_{b}^{*}}$ (left panel) and $\lambda^{*}_{\Sigma_{c}^{*}}/\lambda_{\Sigma_{c}^{*}}$ (right panel) versus  $\rho_N/\rho^{\textrm{sat}}_N$ at central value of Borel mass $M^2$ and different values of continuum threshold.}
\end{figure}
\begin{figure}[h!]
\label{fig5}
\centering
\begin{tabular}{cc}
\epsfig{file=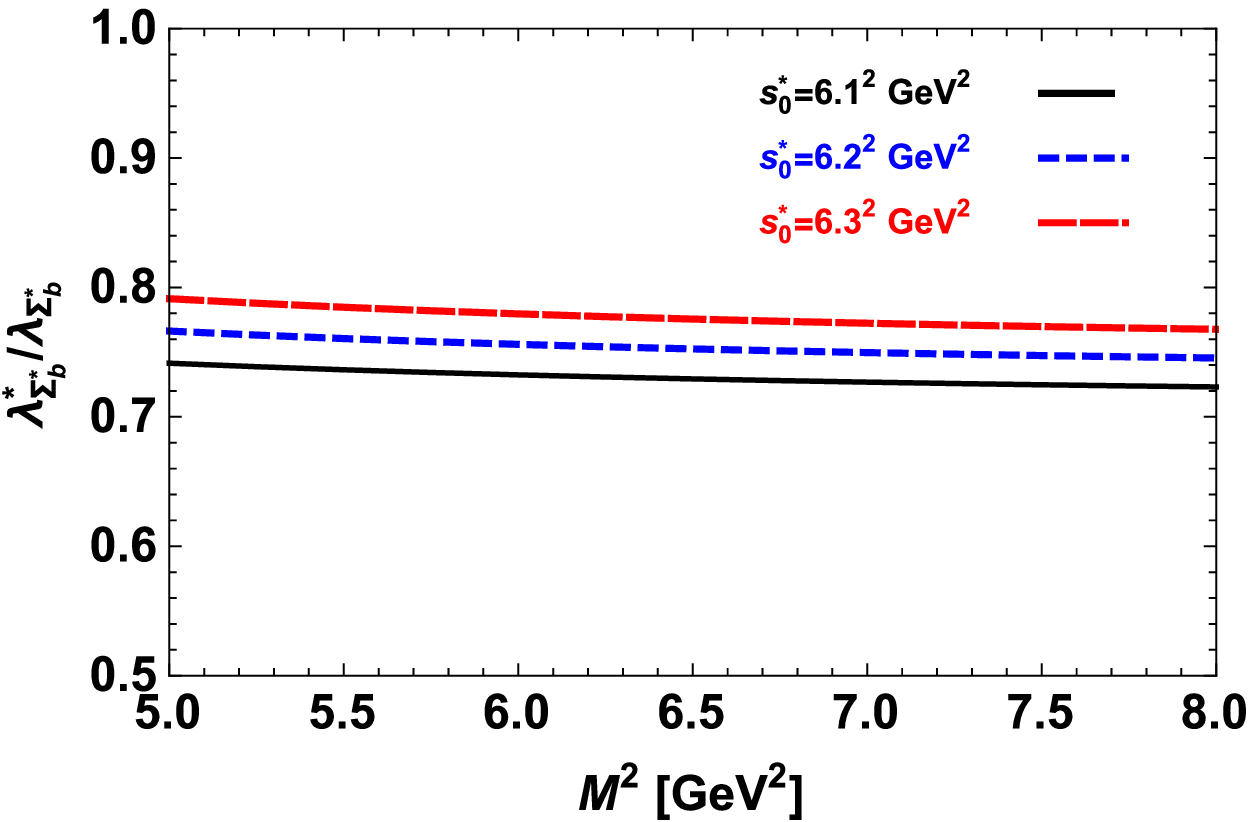,width=0.45\linewidth,clip=} &
\epsfig{file=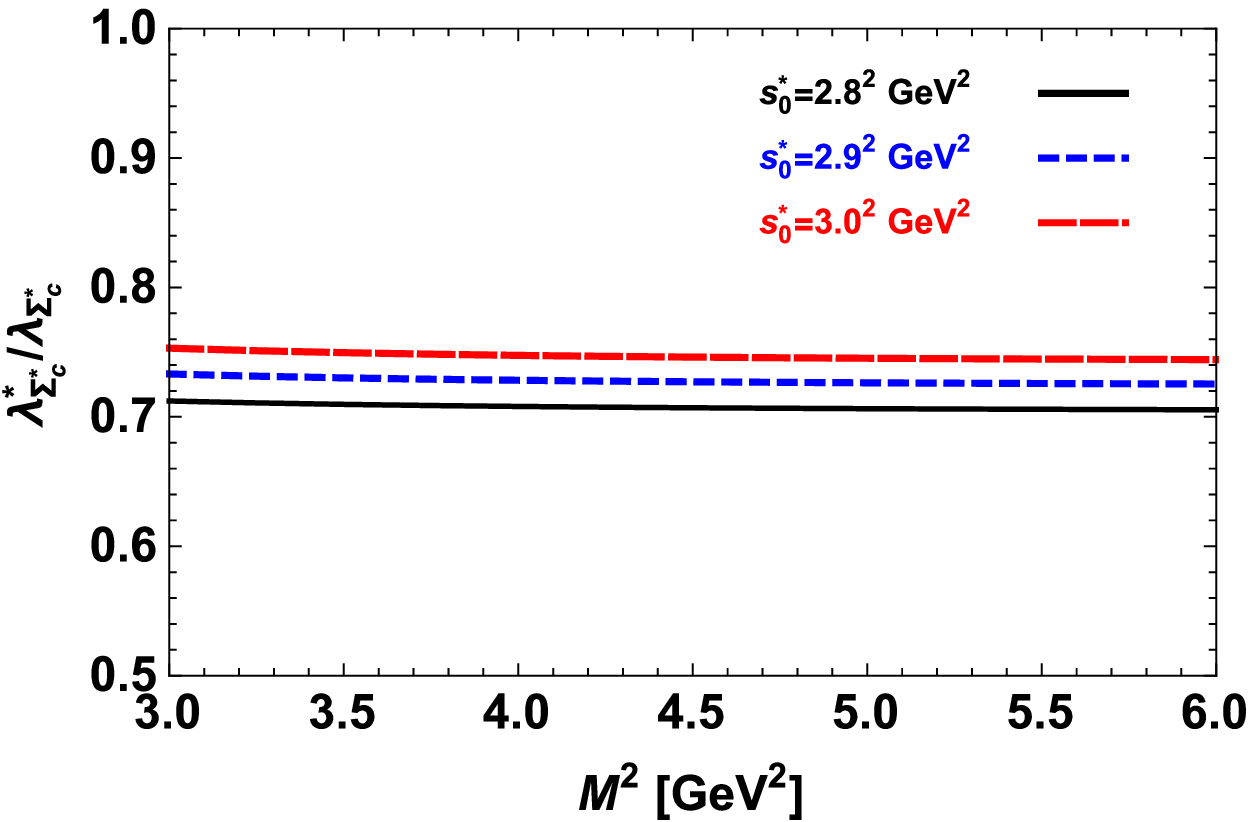,width=0.45\linewidth,clip=}  
\end{tabular}
\caption{The same ratios as figure 5, but as  functions of  $M^2$ at the saturated nuclear matter density.}
\end{figure}
\end{widetext}

The in-medium residue to vacuum residue ratios of $\Sigma_Q^{*}$ baryons as  functions of density and Borel parameter are presented in figures 5 and 6, respectively. The results are presented in three different values of continuum threshold parameter, and at average value of the Borel mass in figure 5 and at saturation nuclear matter density in figure 6. In both channels, the results do not strongly depend on the continuum threshold parameter. It is seen that the quantities under consideration decreases linearly with increasing the medium density and are roughly insensitive  to the changes of the Borel mass parameter. These residue ratios  are depicted for  $\Xi_Q^{*}$ baryons in figures 7 and 8, as well as for $\Omega_Q^{*}$ baryons in figures 9 and 10. We observe similar behaviors in these channels. Among all these channels, at the saturation nuclear  matter density, the maximum shift in the residue ratios is in $\Sigma_c^{*}$ channel and the minimum one belongs to the $\Omega_b^{*}$ baryon.

\begin{widetext}

\begin{figure}[h!]
\label{fig6}
\centering
\begin{tabular}{cc}
\epsfig{file=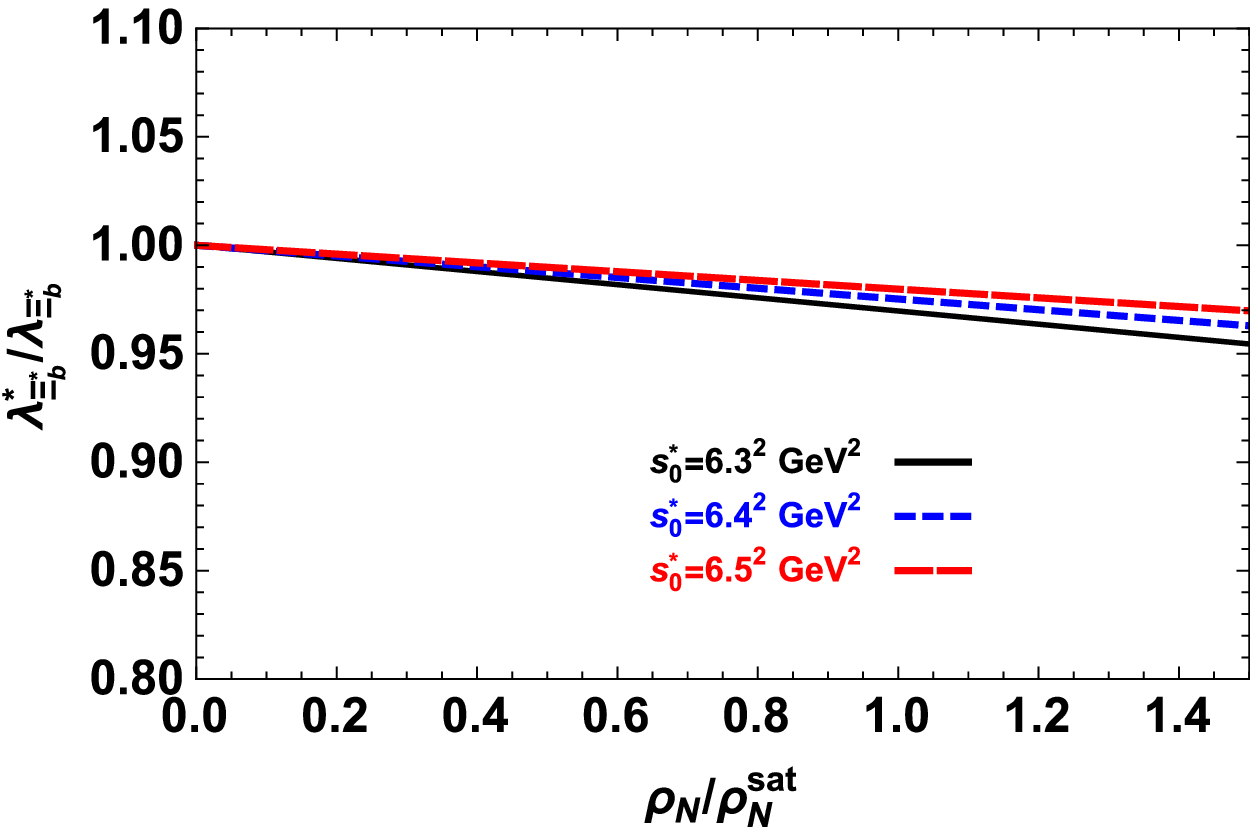,width=0.45\linewidth,clip=} &
\epsfig{file=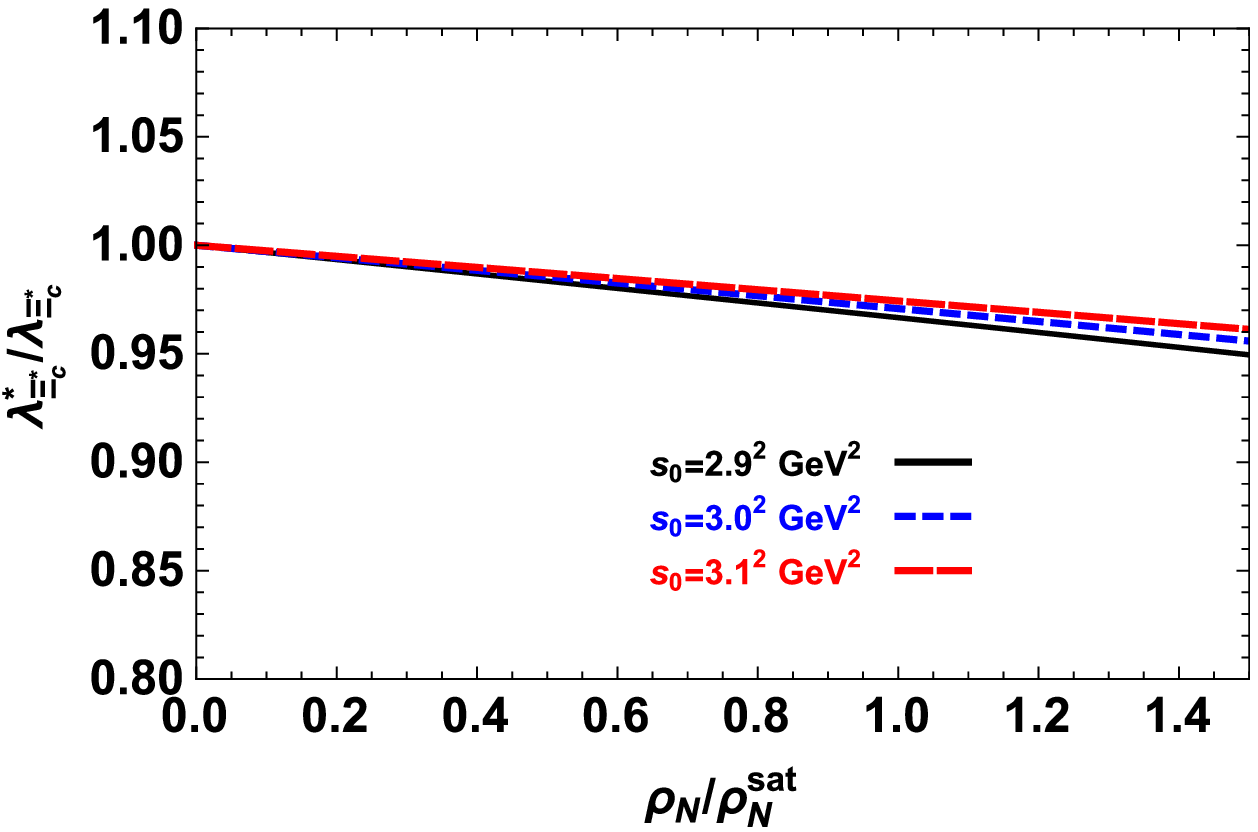,width=0.45\linewidth,clip=}  
\end{tabular}
\caption{The same as figure 5, but for $ \lambda^{*}_{\Xi_{b}^{*}}/\lambda_{\Xi_{b}^{*}}$ (left-panel) and $\lambda^{*}_{\Xi_{c}^{*}}/\lambda_{\Xi_{c}^{*}}$ (right-panel).}
\end{figure}
\begin{figure}[h!]
\label{fig7}
\centering
\begin{tabular}{cc}
\epsfig{file=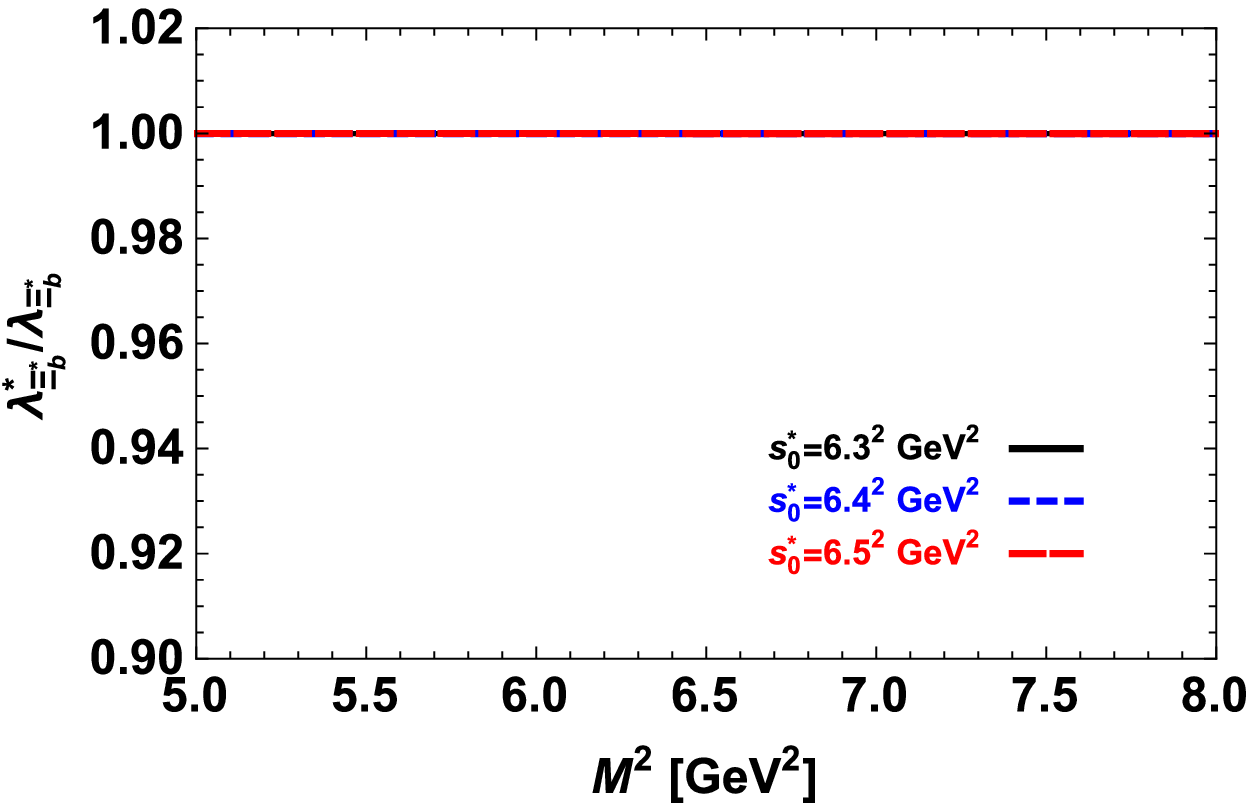,width=0.45\linewidth,clip=} &
\epsfig{file=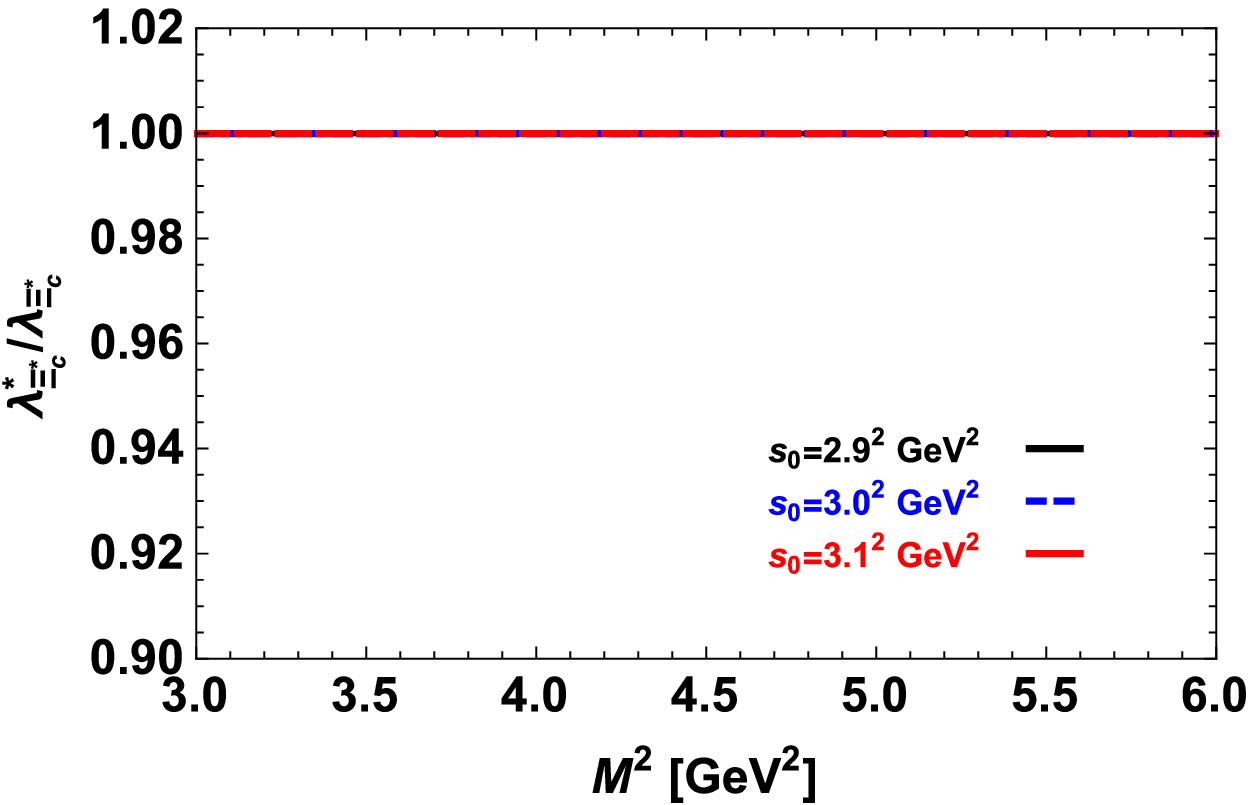,width=0.45\linewidth,clip=}  
\end{tabular}
\caption{The same as figure 6, but for $ \lambda^{*}_{\Xi_{b}^{*}}/\lambda_{\Xi_{b}^{*}}$ (left-panel) and $\lambda^{*}_{\Xi_{c}^{*}}/\lambda_{\Xi_{c}^{*}}$ (right-panel).}
\end{figure}
\end{widetext}

\begin{widetext}

\begin{figure}[h!]
\label{fig8}
\centering
\begin{tabular}{cc}
\epsfig{file=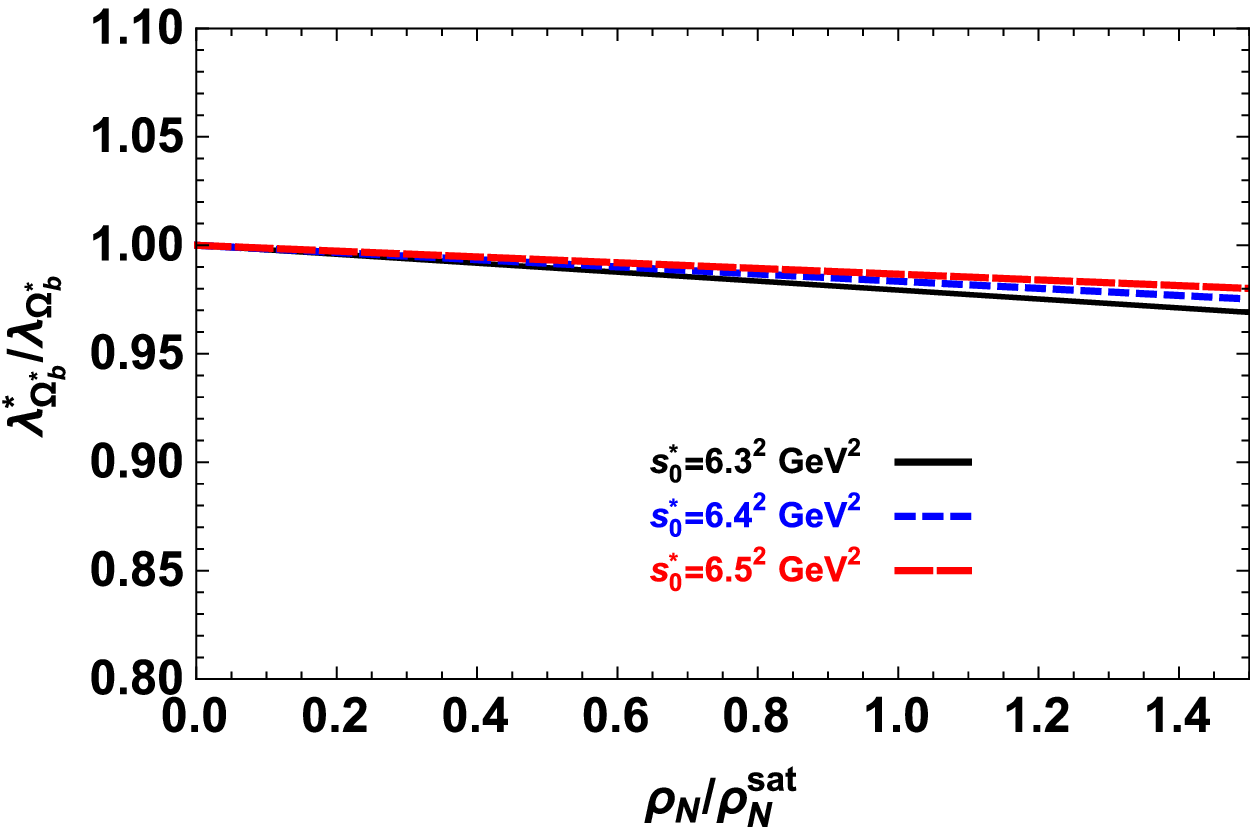,width=0.45\linewidth,clip=} &
\epsfig{file=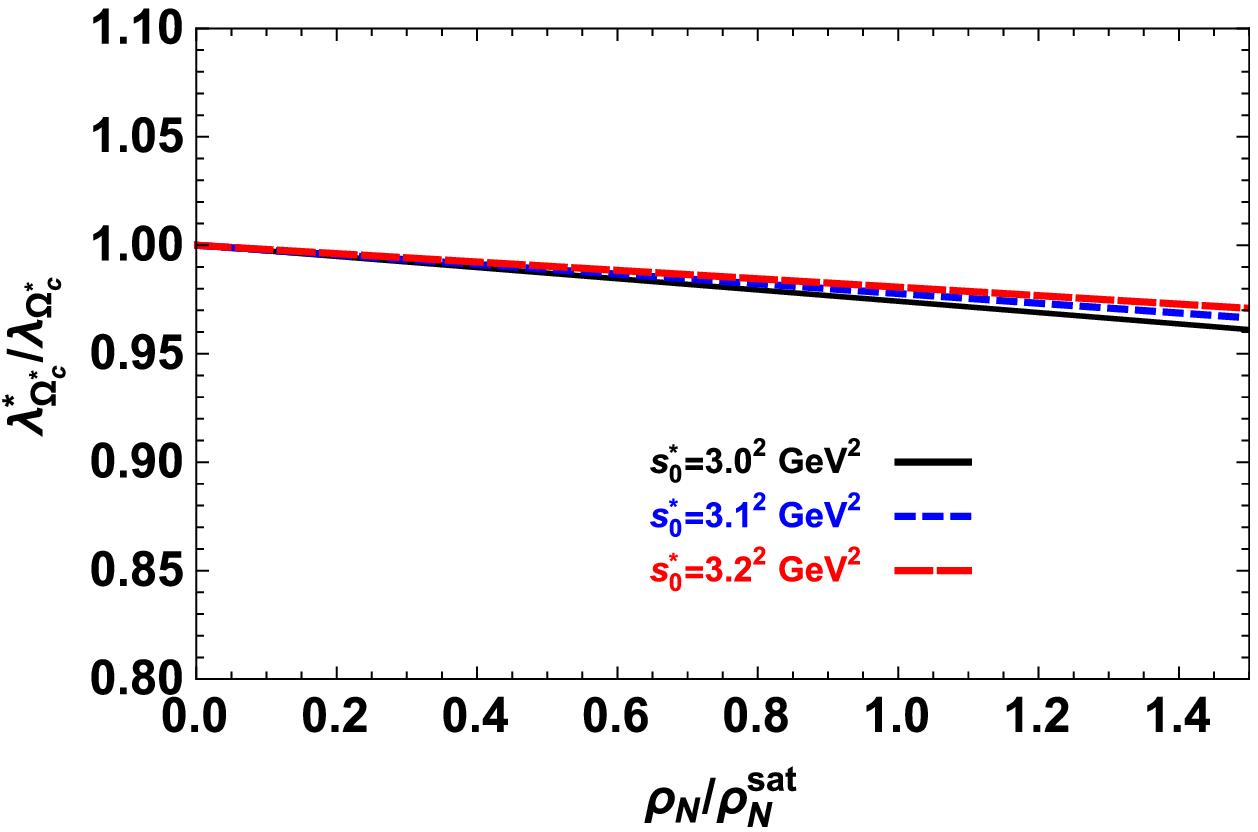,width=0.45\linewidth,clip=}  
\end{tabular}
\caption{The same as figure 5, but for  $ \lambda^{*}_{\Omega_{b}^{*}}/\lambda_{\Omega_{b}^{*}}$ (left-panel) and $\lambda^{*}_{\Omega_{c}^{*}}/\lambda_{\Omega_{c}^{*}}$ (right-panel).}
\end{figure}
\begin{figure}[h!]
\label{fig9}
\centering
\begin{tabular}{cc}
\epsfig{file=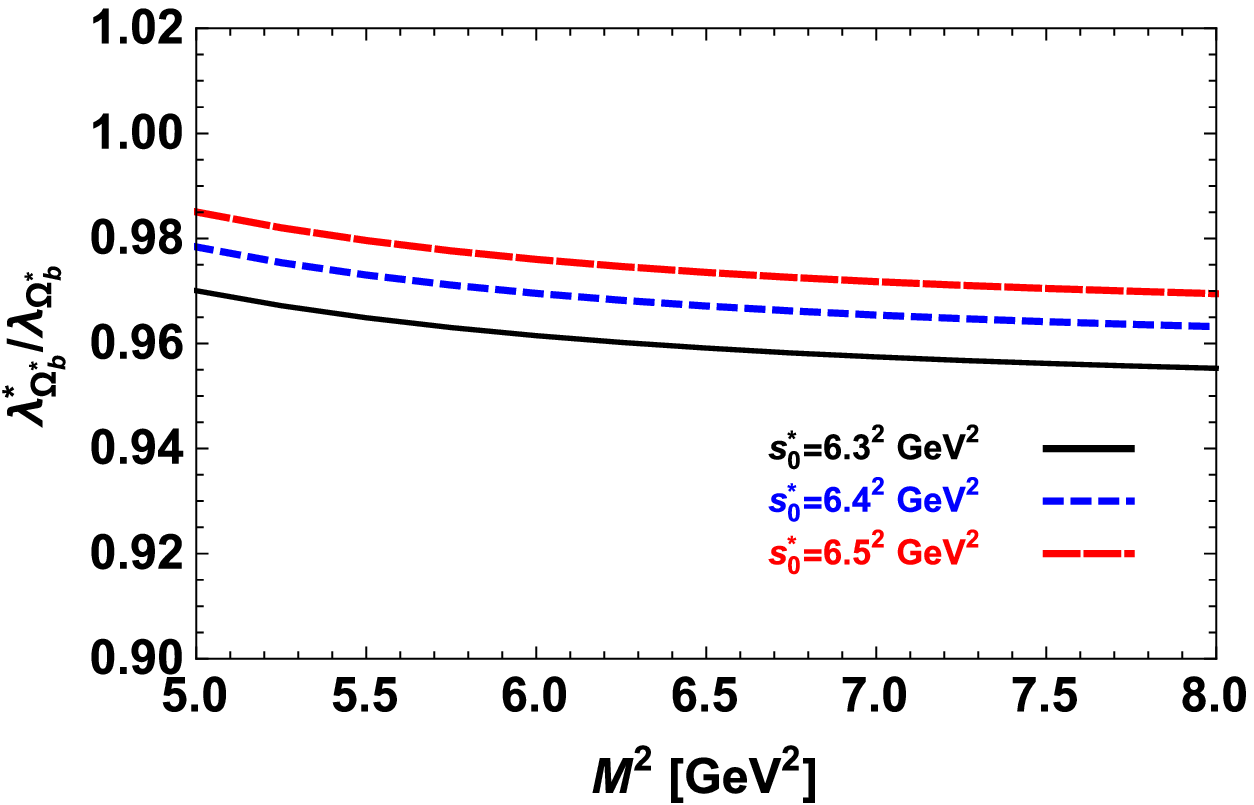,width=0.45\linewidth,clip=} &
\epsfig{file=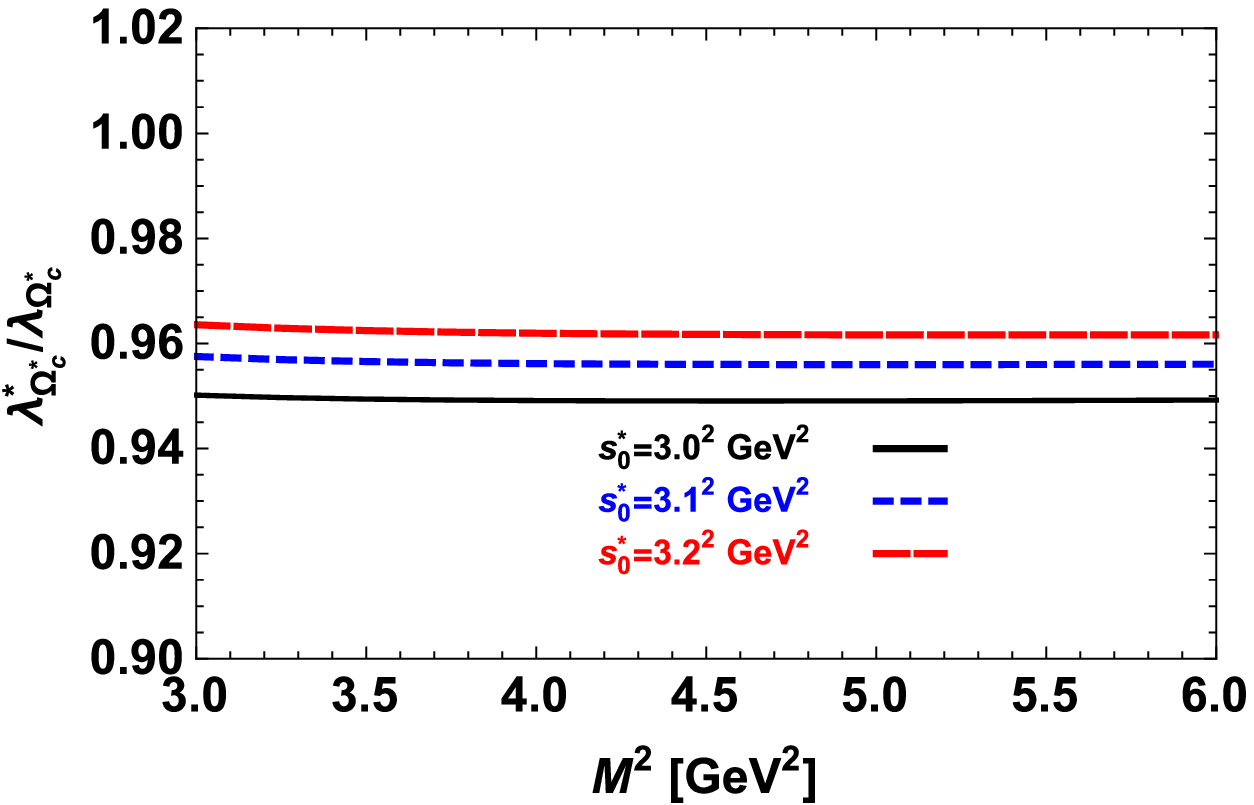,width=0.45\linewidth,clip=}  
\end{tabular}
\caption{The same as figure 6, but for  $ \lambda^{*}_{\Omega_{b}^{*}}/\lambda_{\Omega_{b}^{*}}$ (left-panel) and $\lambda^{*}_{\Omega_{c}^{*}}/\lambda_{\Omega_{c}^{*}}$ (right-panel).}
\end{figure}
\end{widetext}

Obtained from  analyses, we collect the average values of the ratios $m^{*}_{B_{Q}}/m_{B_{Q}}$,  $\Sigma_{v,B_{Q}}/m_{B_{Q}}$ and  $\lambda^{*}_{B_{Q}}/\lambda_{B_{Q}}$ at saturation density in Table III. Among the presented results, we see considerable negative shifts  in the masses of the $\Sigma_b^{*}$ and $\Sigma_c^{*}$ baryons, representing the attraction of these states by the medium. The masses of $\Xi_Q^{*}$ and   $\Omega_Q^{*}$ baryons remains roughly unaffected by the medium. As far as the ratios of the residues are concerned, we see again  considerable negative shifts in $\Sigma_Q^{*}$ channels. The shift in the residue of $\Sigma_c^{*}$ baryon is maximum and reaches to roughly $  27\% $ of its  vacuum value. The residues of  $\Xi_Q^{*}$ and   $\Omega_Q^{*}$ baryons are slightly affected by the medium. The maximum vector self energy belongs to the $\Sigma_b^{*}$ baryon in cold nuclear matter and amounts $ 420 $ $ MeV $, while the minimum one belongs to the $\Omega_c^{*}$ baryon with amount of  $ 55 $ $ MeV $. 
\begin{table}[ht!]
\centering
\begin{tabular}{ |l|c|c|c|}
\hline \hline
&$m^{*}_{B_{Q}}/m_{B_{Q}}$  &  $\Sigma_{v,B_{Q}}/m_{B_{Q}}$ & $\lambda^{*}_{B_{Q}}/\lambda_{B_{Q}}$  \\ \hline
$\Sigma_b^{*}$& $0.946^{+0.015}_{-0.018}$ & $0.072^{+0.017}_{-0.015}$ & $0.755^{+0.037}_{-0.032}$\\ 
$\Sigma_c^{*}$& $0.985^{+0.011}_{-0.010}$ & $0.054^{+0.010}_{-0.009}$ & $0.728^{+0.025}_{-0.023}$\\ 
$\Xi_b^{*}$& $0.999^{+0.011}_{-0.009}$ & $0.005^{+0.001}_{-0.001}$ & $0.976^{+0.009}_{-0.008}$\\ 
$\Xi_c^{*}$& $1.003^{+0.001}_{-0.001} $& $0.004^{+0.001}_{-0.001}$ & $0.971^{+0.005}_{-0.004}$\\ 
$\Omega_b^{*}$& $1.004^{+0.002}_{-0.001}$ & $0.002^{+0.001}_{-0.001}$ & $0.984^{+0.008}_{-0.007}$\\
$\Omega_c^{*}$& $1.007^{+0.003}_{-0.002}$ & $0.002^{+0.001}_{-0.001}$ & $0.978^{+0.004}_{-0.003}$\\ \hline
\end{tabular}
\caption{Mean values of the ratios $m^{*}_{B_{Q}}/m_{B_{Q}}$,   $\Sigma_{v,B_{Q}}/m_{B_{Q}}$ and  $\lambda^{*}_{B_{Q}}/\lambda_{B_{Q}}$ at saturation nuclear matter density. }
\end{table}


In summary, the in-medium properties of the spin$-3/2$  $\Sigma_Q^{*}$, $\Xi_Q^{*}$ and   $\Omega_Q^{*}$ heavy baryons have been investigated using the  in-medium  two point QCD sum rule method. We discussed the behaviors of the ratios $m^{*}_{B_{Q}}/m_{B_{Q}}$,  $\Sigma_{v,B_{Q}}/m_{B_{Q}}$ and  $\lambda^{*}_{B_{Q}}/\lambda_{B_{Q}}$ with respect to the changes of the cold nuclear matter density as well as the two auxiliary parameters entered the calculations. We observed that the masses and residues of the $\Sigma_Q^{*}$ baryons receive considerable shifts due to the medium, representing attractions of these states by the nuclear matter. The masses and residues of the   $\Xi_Q^{*}$ and   $\Omega_Q^{*}$ baryons are not affected by the medium, considerably.    The maximum vector self energy belongs to the $\Sigma_b^{*}$ baryon in cold nuclear matter with amount of $ 420 $ $ MeV $, while the minimum one belongs to the $\Omega_c^{*}$ channel, which amounts  $ 55 $ $ MeV $. The results of the present study may be used in analyses of the results of the heavy ion collision experiments as well as in the study of the heavy baryon-nucleon interactions.
Some in-medium experiments like PANDA plan to measure parameters of the heavy hadrons, especially the charmed  channels,  in nuclear matter (for instance see Refs. \cite{Wolf:2017jbv,Wolf:2017qbz}). Our predictions can also be used in analyses of the results of such experiments.

\section{Acknowledgment}
K.~A. thanks Do\v{g}u\c{s} University for the partial financial support
through the grant BAP 2015-16-D1-B04.


\appendix*
\section{Correlation functions of OPE side in terms of quark propagators for $\Sigma_Q^{*}, \Xi_Q^{*}$ and $\Omega_Q^{*}$ baryons}
In this Appendix, we collect the results for correlation functions in OPE side obtained after inserting the interpolating currents and contracting out the quark fields. For the particles $\Sigma_Q^{*}$ and $\Xi_Q^{*}$ containing two different light quarks, we get
\begin{widetext}
\begin{eqnarray}
\Pi_{\mu\nu}^{\Sigma_Q^{*},\Xi_Q^{*}} =& -& \frac{2}{3}i\epsilon_{abc}\epsilon_{a'b'c'}\int d^4 x e^{ipx} \Big\langle \psi_0\Big| \Big \{ S^{ca'}_{Q}\gamma_{\nu}S'^{bb'}_{q_2}\gamma_{\mu}S^{ac'}_{q_1}+S^{cb'}_{Q}\gamma_{\nu}S'^{aa'}_{q_1}\gamma_{\mu}S^{bc'}_{q_2} +S^{ca'}_{q_2}\gamma_{\nu}S'^{bb'}_{q_1}\gamma_{\mu}S^{ac'}_{Q}  \nonumber \\
&+& S^{cb'}_{q_2}\gamma_{\nu}S'^{aa'}_{Q}\gamma_{\mu}S^{bc'}_{q_1} + S^{cb'}_{q_1}\gamma_{\nu}S'^{aa'}_{q_2}\gamma_{\mu}S^{bc'}_{Q} + S^{ca'}_{q_1}\gamma_{\nu}S'^{bb'}_{Q}\gamma_{\mu}S^{ac'}_{q_2} + Tr\big [ \gamma_{\mu}S^{ab'}_{q_1} \gamma_{\nu} S'^{ba'}_{q_2} \big] S^{cc'}_{Q}\nonumber \\
&+&  Tr\big [ \gamma_{\mu}S^{ab'}_{Q} \gamma_{\nu} S'^{ba'}_{q_1} \big] S^{cc'}_{q_2} + Tr\big [ \gamma_{\mu}S^{ab'}_{q_2} \gamma_{\nu} S'^{ba'}_{Q} \big] S^{cc'}_{q_2}\Big| \psi_0\Big\rangle, \nonumber \\
\end{eqnarray}
\end{widetext}
where  the light quark flavours $q_1$ and $q_2$ are given in Table \ref{tab:QF}. For $\Omega_Q^{*}$,  we have
\begin{widetext}
\begin{eqnarray}
\Pi_{\mu\nu}^{\Omega_Q^{*}} =& -& \frac{1}{3}i\epsilon_{abc}\epsilon_{a'b'c'}\int d^4 x e^{ipx} \Big\langle \psi_0\Big| \Big \{ S^{ca'}_{Q}\gamma_{\nu}S'^{ab'}_{s}\gamma_{\mu}S^{bc'}_{s} - S^{ca'}_{Q}\gamma_{\nu}S'^{bb'}_{s}\gamma_{\mu}S^{ac'}_{s} - S^{cb'}_{Q}\gamma_{\nu}S'^{aa'}_{s}\gamma_{\mu}S^{bc'}_{s} \nonumber \\
&+& S^{cb'}_{Q}\gamma_{\nu}S'^{ba'}_{s}\gamma_{\mu}S^{ac'}_{s} + S^{ca'}_{s}\gamma_{\nu}S'^{ab'}_{Q}\gamma_{\mu}S^{bc'}_{s} - S^{ca'}_{s}\gamma_{\nu}S'^{bb'}_{Q}\gamma_{\mu}S^{ac'}_{s} + S^{ca'}_{s}\gamma_{\nu}S'^{ab'}_{s}\gamma_{\mu}S^{bc'}_{Q} \nonumber \\
&-& S^{ca'}_{s}\gamma_{\nu}S'^{bb'}_{s}\gamma_{\mu}S^{ac'}_{Q} - S^{cb'}_{s}\gamma_{\nu}S'^{aa'}_{Q}\gamma_{\mu}S^{bc'}_{s} + S^{cb'}_{s}\gamma_{\nu}S'^{ba'}_{Q}\gamma_{\mu}S^{ac'}_{s} - S^{cb'}_{s}\gamma_{\nu}S'^{aa'}_{s}\gamma_{\mu}S^{bc'}_{Q} \nonumber \\
&+&S^{cb'}_{s}\gamma_{\nu}S'^{ba'}_{s}\gamma_{\mu}S^{ac'}_{Q} - S^{cc'}_{s}Tr\big[ S^{ba'}_{Q}\gamma_{\nu}S'^{ab'}_{s}\gamma_{\mu}\big] + S^{cc'}_{s}Tr\big[ S^{bb'}_{Q}\gamma_{\nu}S'^{aa'}_{s}\gamma_{\mu}\big] - S^{cc'}_{s}Tr\big[ S^{ba'}_{s}\gamma_{\nu}S'^{ab'}_{Q}\gamma_{\mu}\big] \nonumber \\
&-& S^{cc'}_{Q}Tr\big[ S^{ba'}_{s}\gamma_{\nu}S'^{ab'}_{s}\gamma_{\mu}\big] + S^{cc'}_{s}Tr\big[ S^{bb'}_{s}\gamma_{\nu}S'^{aa'}_{Q}\gamma_{\mu}\big] + S^{cc'}_{Q}Tr\big[ S^{bb'}_{s}\gamma_{\nu}S'^{aa'}_{s}\gamma_{\mu}\big]\Big| \psi_0\Big\rangle. \nonumber \\
\end{eqnarray}
\end{widetext}
where $S'^{ij}_{n}=CS'^{ij,T}_{n}C$. The abbrevation $Tr[...]$ represents the trace of the gamma matrices. In the fixed point gauge, the expressions of light $S_q$ and heavy $S_Q$ quark propagators are given in Refs. \cite{Cohen:1994wm,Azizi:2016dmr}.


\end{document}